\documentclass[pra,onecolumn,superscriptaddress,aps,5pt]{revtex4}
\usepackage{amsfonts}
\usepackage{amssymb,latexsym,amsmath}
\usepackage{graphicx}
\usepackage{float}
\usepackage{dcolumn}
\usepackage{times}
\usepackage{subfigure}
\usepackage[toc,page,title,titletoc,header]{appendix}
\usepackage[usenames]{color}
\usepackage{ulem}

\begin{document}

\title{Generalized Dimers and their Stokes-variable Dynamics}
\author{H. Xu}
\affiliation{Department of Mathematics and Statistics, University of Massachusetts,
Amherst MA 01003-4515, USA}
\author{P. G.\ Kevrekidis }
\affiliation{Department of Mathematics and Statistics, University of Massachusetts,
Amherst MA 01003-4515, USA}
\author{A. Saxena}
\affiliation{Center for Nonlinear Studies and Theoretical Division, Los Alamos National Laboratory, Los Alamos, New Mexico 87545, USA}

\begin{abstract}
In the present work, we generalize the setting of dimers with potential
gain and loss which have been extensively considered recently in
$\mathcal{P T}$-symmetric contexts. We consider a pair of waveguides
which are evanescently coupled but may also be actively coupled
and may possess onsite gain and loss, as well as (possibly non-uniform)
nonlinearity. We identify (and where appropriate review
from earlier work) a plethora of interesting dynamical scenaria ranging
from the existence of stable and unstable fixed points and integrable
dynamics, to the emergence of pitchfork or Hopf bifurcations and the generation
of additional fixed points and limit cycles, respectively, as well as
the potential deviation of trajectories to infinity. Thus, a catalogue
of a large number of possible cases is given and their respective
settings physically justified (where appropriate).
\end{abstract}

\maketitle

\section{Introduction}

Over the past decade and a half, there has been an intense interest
in the theme of open systems featuring gain and loss due to numerous
developments in the study of $\mathcal{PT}$-symmetric dynamics; see
e.g. \cite{Bender_review,special-issues,review}.
While the original proposal of such systems was given in the context
of quantum mechanical (non-Hermitian, yet still potentially
bearing real eigenvalues) Hamiltonians, relevant applications
sprang in a number of diverse areas of physical interest.
In particular, the analogy between the Schr\"{o}dinger
equation in quantum mechanics and the paraxial propagation equation in
optics led to such proposals in optics~\cite{Muga,PT_periodic}
which were subsequently realized in a series of experiments~\cite{experiment}.
Additionally,  ``engineered''
$\mathcal{PT}$ symmetric systems  also arose in the context
of electronic circuits; see the work
of~\cite{tsampikos_recent} and also the review of~\cite{tsampikos_review}.
Further developments including mechanical systems~\cite{pt_mech} and even
whispering-gallery microcavities~\cite{pt_whisper} have recently followed.

 The realization of $\mathcal{P T}$ symmetry in optical settings
naturally brought forth the question of the implications of
nonlinearity in such systems, as nonlinearity is rather ubiquitous
in optics~\cite{Musslimani2008}. This, in turn, led to the exploration
of structures such as bright~\cite{Yang} and dark~\cite{dark,vortices}
solitons, two-dimensional generalizations of solitons~\cite{midya},
as well as vortices~\cite{vortices}. Generalizations of
one-dimensional~\cite{dual,dual2} and
two-dimensional~\cite{Burlak} such entities were predicted to
exist as stable objects in dual-core couplers (with the gain
and loss being present in different cores). In addition,
a large number of studies focused on the context of discrete systems
in the setting of few sites (so-called oligomers), as well
as lattices~\cite{kot1,sukh1,kot2,grae1,grae2,pgk,dmitriev1,dmitriev2,R30add5,R34,R44,R46,baras1,baras2,konorecent3,leykam,konous,uwe,tyugin,pickton}.

The present work is in the spirit of oligomers and more
specifically of dimers examined rather exhaustively in recent
years both in experiments~\cite{experiment} and in
theory~\cite{kot1,sukh1,grae1,grae2,pgk,R44,R46,tyugin,pickton}.
However, it also fundamentally differs from these works, as it is
{\it not} in the ``standard'' (and rather delicate in its
necessitated balance of gain and loss) $\mathcal{PT}$-symmetric realm.
Instead, we seek to further explore a setting put forth in the
recent, fundamental contribution of~\cite{barasflach}, where
an active medium with two waveguides, each contributing to the
gain of the other is set forth, coupled with the intrinsic loss
of each waveguide. While the resulting setting of~\cite{barasflach}
is not~$\mathcal{PT}$-symmetric, it is demonstrated that it
features a robust feedback loop, which, in turn, generates
stationary and oscillatory regimes in a wide range of gain-loss
parameter values. Additionally, the resulting system as written
in the so-called Stokes variables presents an intriguing proximity
to the classical Lorenz dynamical system which is a prototype
of chaotic dynamics. Indeed, the work of~\cite{barasflach}
revealed a sizable region of their two dimensional (gain-loss)
parameter space exhibiting chaotic dynamics.

Our aim here is to present a general form of the relevant
dimer system inter-twining the characteristics of
(a) an active medium, (b) evanescent coupling, (c) intrinsic loss or
imposed gain on each waveguide and finally nonlinearity (of possibly
even non-uniform type) and to explore the situation for different
types of combinations of these characteristics. In each of the
cases considered, we explore the Stokes variable formalism
and rewrite the dynamical system as a 3 degree-of-freedom
setup (effectively removing an overall trivial phase, associated
with the U$(1)$ invariance of the dimer). Subsequently, we
perform the ``local'' analysis of the system, identifying its
fixed points and their spectral characteristics, as well elucidating
the potential bifurcations (pitchforks, Hopfs, etc.). Finally,
we complement the theoretical analysis with direct numerical computations
which illustrate prototypical dynamics of the respective cases.

The presentation is structured as follows. In section II, we briefly
present the general setup associated with our system. In section III,
we catalogue the relevant subcases and their theoretical (fixed point,
stability and bifurcation) analysis, as well as the associated numerical
computations. Finally, in section IV, we summarize our main findings
and present a number of conclusions and directions for future work.

\section{Theoretical Setup}

We start by considering in the spirit of~\cite{barasflach}
the general dimer system for the evolution of two
waveguides with  variables $u_1(t)$, $u_2(t)$ according to:
\begin{eqnarray}
\dot{u}_1&=&(a_1+i\sigma)u_2+b_1 u_1-i P_1|u_1|^2u_1,\\
\dot{u}_2&=&(a_2+i\sigma)u_1+b_2 u_2-i P_2|u_2|^2u_2.
\label{eqn1}
\end{eqnarray}
Here, $a_{1,2}$ play the role of the active underlying medium, which
we will broadly conceive of as not necessarily symmetrically acting
on the two waveguides (although the latter is the canonical physical
case). Here, for mathematical purposes, we will also expand the realm
of considerations to cases with $a_1 \neq a_2$. On the other hand,
$\sigma$ plays the role of the evanescent coupling between the waveguides
and $b_{1,2}$ is the respective gain and loss imposed individually
(or intrinsically) on each waveguide. Finally, $P_{1,2}$ will represent
the nonlinearity strength of the cubic Kerr effect. For the latter,
again motivated
by physical cases of different material properties as e.g. in
settings of the form of~\cite{towers,martincent}, we will not a priori
assume equal strength of the nonlinear prefactors.

In the following discussion, we will use the well-established
Stokes variables~\cite{kot1,tyugin,pickton,barasflach}
$X=\frac{1}{2}(|u_1|^2-|u_2|^2)$, $Y=\frac{i}{2}(u_1u_2^*-u_2u_1^*)$ and
$Z=\frac{1}{2}(u_1u_2^*+u_2u_1^*)$. By also rescaling time as
$\tilde{t}=2t$, our dynamical equations can be rewritten as follows:
\begin{eqnarray}
\frac{dX}{d\tilde{t}}&=&\frac{a_1-a_2}{2}Z-\sigma Y+\frac{1}{2}(b_1|u_1|^2-b_2|u_2|^2),\\
\frac{dY}{d\tilde{t}}&=&\frac{b_1+b_2}{2}Y+\sigma X+\frac{1}{2}(P_1|u_1|^2-P_2|u_2|^2)Z,\\
\frac{dZ}{d\tilde{t}}&=&\frac{1}{2}(a_2|u_1|^2+a_1|u_2|^2)+\frac{b_1+b_2}{2}Z-(P_1|u_1|^2-P_2|u_2|^2)Y.
\label{eqn2}
\end{eqnarray}
We now let $N=\frac{1}{2}(|u_1|^2+|u_2|^2)$ i.e., the
optical power, and thus $X^2+Y^2+Z^2=N^2$; notice that
given this formula and the definition of $X$ and $N$, we can always
express the above equations as a three-dimensional dynamical system.
The differential equation for the evolution of $N$ reads:
\begin{eqnarray}
\frac{dN}{d\tilde{t}}&=&\frac{a_1+a_2}{2}Z+\frac{1}{2}(b_1|u_1|^2+b_2|u_2|^2).
\label{eqn2_00}
\end{eqnarray}
With conditions $a_1+a_2=0$ and $b_1=b_2$, $N$ is exponentially
increasing ($b>0$) or decreasing ($b<0$), hence it is straightforward to
establish instability $(b>0)$ or stability $(b<0)$ about the origin, which is
the only fixed point. Hence we focus more on systems where parameters do not
satisfy these conditions.

It is also interesting to note that if we write $u_1=\rho_1e^{i(\phi+\theta)}$ and $u_2=\rho_2e^{i\phi}$, then the Stokes Variables can be expressed using
$\rho_1$, $\rho_2$ and $\theta$, i.e., the overall free phase due to the
gauge invariance of the model is the dynamical variable that has been
eliminated in this system.
More specifically, $X=\frac{1}{2}(\rho_1^2-\rho_2^2)$,
$Y=\rho_1\rho_2 \sin\theta$ and $Z=\rho_1\rho_2 \cos\theta$. Based only on these
three equations, $Y$ and $Z$ are two similar variables (playing the role
of polar coordinates for the pair $(\rho_1 \rho_2, \theta)$,
while $X$ measures the intensity difference between the waveguides.

Reversing the relevant transformation, $\rho_1$ and $\rho_2$ and $\theta$ can
also be obtained from Stokes Variables: $\theta={\rm arctan}\frac{Y}{Z}$, $\rho_1^2=\sqrt{X^2+Y^2+Z^2}+X$ and $\rho_2^2=\sqrt{X^2+Y^2+Z^2}-X$.
In what follows, we will solely consider the analysis of the three-dimensional
dynamical system in the realm of Stokes Variables, bearing the above
transformations and inverse transformations in mind.

\section{Catalogue of Different Parametric Cases}

\subsection{$a_1=a_2=0$}

This is the case connecting to the earlier work including
in $\mathcal{P T}$-symmetric settings where the medium is not active in
that it does not induce a gain or loss related coupling between
the waveguides.

\subsubsection{$b_1=b_2=b$}
Since $a_1=a_2=0$ automatically leads to $a_1+a_2=0$, if we also assume equal gain or loss on two waveguides ($b_1=b_2=b$), the system is quite straightforward: the origin is the sole fixed point and it is either a
stable spiral (for $b < 0$) or an unstable spiral (for $b > 0$). Also note that the Stokes variable system bears an obvious reflection symmetry if $(x, y)$ is replaced by $(-x,-y)$ (for $P_1=P_2$) or replacing $(x , y , z)$ by $(-x, -y, -z)$ (for $P_1=-P_2$).

\subsubsection{$b_1=-b_2=b$, $P_1=P_2=P$}
The system in this case reads:
\begin{eqnarray}
\frac{dX}{d\tilde{t}}&=&-\sigma Y+ b N,
\label{eqn2z}
\\
\frac{dY}{d\tilde{t}}&=&\sigma X+ P X Z,
\label{eqn2a}
\\
\frac{dZ}{d\tilde{t}}&=&-P X Y.
\label{eqn2b}
\end{eqnarray}
In fact, this is the well studied case of
the $\mathcal{P T}$-symmetric dimer. For the latter, it is
well known that it possesses two conserved quantities $C^2=Y^2+(Z+\frac{\sigma}{P})^2$ and $J=N+\frac{b}{p} \arcsin(\frac{Z+\frac{\sigma}{P}}{C})$,
and is hence integrable, possessing also the potential for periodic
orbits. The ability of the system depending on the relative values
of $b$ and $\sigma$ to settle into periodic orbits or to escape
to $\infty$ has been elucidated in a series of recent
works~~\cite{kot1,tyugin,pickton,barasflach2}.

In this case, the differential equation $\frac{d}{d\tilde{t}}(N)=b X$ does not provide direct dynamical
information but solely an a priori bound to the growth of the
system~\cite{tyugin}.
To better understand the dynamics, we reshape Eqs.~(\ref{eqn2a})-(\ref{eqn2b})
into a harmonic oscillator form~\cite{pelinovsky} using the variable $s(\tilde{t})=\int_0^{\tilde{t}}X(t') dt'$. Then the solutions of $Y$ and $Z$ are:
\begin{eqnarray}
Y(\tilde{t})&=&C_1 \cos(P s(\tilde{t}))+C_2 \sin(P s(\tilde{t})),\\
Z(\tilde{t})&=&-\frac{\sigma}{P}+C_2 \cos(P s(\tilde{t}))-C_1 \sin(P s(\tilde{t})) ,
\label{eqn2b2}
\end{eqnarray}
where $C_1$ and $C_2$ are constants, determined by the initial conditions. These imply not only that $Y$ and $Z$ are always bounded, but also
that $Y^2+(Z+\frac{\sigma}{P})^2$ is constant over time. Thus,
the dynamical system is actually two dimensional and any trajectory is
constrained on the surface of a cylinder. By exploring the dynamics of $|u_1|^2$ and $|u_2|^2$, we can show that the density can become
unbounded if the initial values are suitably chosen. Besides,
an example of explicit unbounded solution has been reported
in~\cite{barasflach2}.

The fixed points in this case are $(0,0,0)$ and
$(0, \frac{b}{\sigma}N, \pm\frac{\sqrt{\sigma^2-b^2}}{\sigma}N)$ for any
non-negative number $N$.
Since $N=\sqrt{X^2+Y^2+Z^2}$ is not continuously differentiable at $(0,0,0)$,
interestingly, it is not straightforward to examine the Jacobian matrix
of the fixed point of the origin (at the level of our Stokes
variable equations). Instead, in such a case it is advisable
to return to the original dynamical system revealing that for $b^2< \sigma^2$
the origin is a stable fixed point (center), while the converse is true
for $b^2> \sigma^2$ (saddle). The two additional
families
of fixed points correspond to the symmetric and anti-symmetric (for $b=0$)
states of the Hamiltonian
analog of the dimer which indeed collide and disappear in a saddle-center
bifurcation at $\sigma^2=b^2$, as is well-known~\cite{pgk} for such
~$\mathcal{P T}$-symmetric dimers.

More generally, in this setting for $b^2< \sigma^2$, as is well-known,
the trajectories may be bounded (periodic) or
unbounded~\cite{kot1,pgk,tyugin,pickton,barasflach2}, while for
$b^2> \sigma^2$, they will be generically unbounded.




\subsubsection{$b_1=-b_2=b$, $P_1=-P_2=P$}
This is an interesting case that to the best of our knowledge
has not been explored previously. Here the linear part of the system
is $\mathcal{P T}$-symmetric, while the nonlinear part is {\it not}.
The dynamical equations for the Stokes variables  in this case read:
\begin{eqnarray}
\frac{dX}{d\tilde{t}}&=&-\sigma Y+ b N,\\
\frac{dY}{d\tilde{t}}&=&\sigma X+ P N Z,\\
\frac{dZ}{d\tilde{t}}&=&-P N Y.
\label{eqn2c}
\end{eqnarray}
Once again, the evolution of the power does not lead to unambiguous results
since
$\frac{d}{d\tilde{t}}(N)=b X$, but solely to an a priori bound for
its maximal growth rate since $X \leq N$ (and hence the maximal growth rate
of the exponential growth of $N$ is $b$).

The origin is the sole fixed point in this case. The linearization around it is again not feasible at the
level of Eq.~(\ref{eqn2c}) but can instead be realized at the level of
Eq.~(\ref{eqn1}) suggesting that the origin is a center for $b^2 < \sigma^2$ and a saddle point for $b^2 > \sigma^2$. It is worth noting that the center here is no longer a center for a pure two-dimensional system, as was the case
in the previous subsection and it, thus,
does not necessarily imply the existence of a periodic orbit.
A complementary perspective of this from the point of view of
the Stokes variables is that when $b^2>\sigma^2$, $|bN|>|\sigma Y|$ so that
$X$ is always increasing or decreasing. Notice that the same inequality
can be used to infer the indefinite growth at the broken
$\mathcal{P T}$-symmetry regime (past the $\mathcal{P T}$-phase transition)
in the context of Eq.~(\ref{eqn2b}).

\begin{figure}[!htbp]
\begin{tabular}{cc}
\includegraphics[width=8cm]{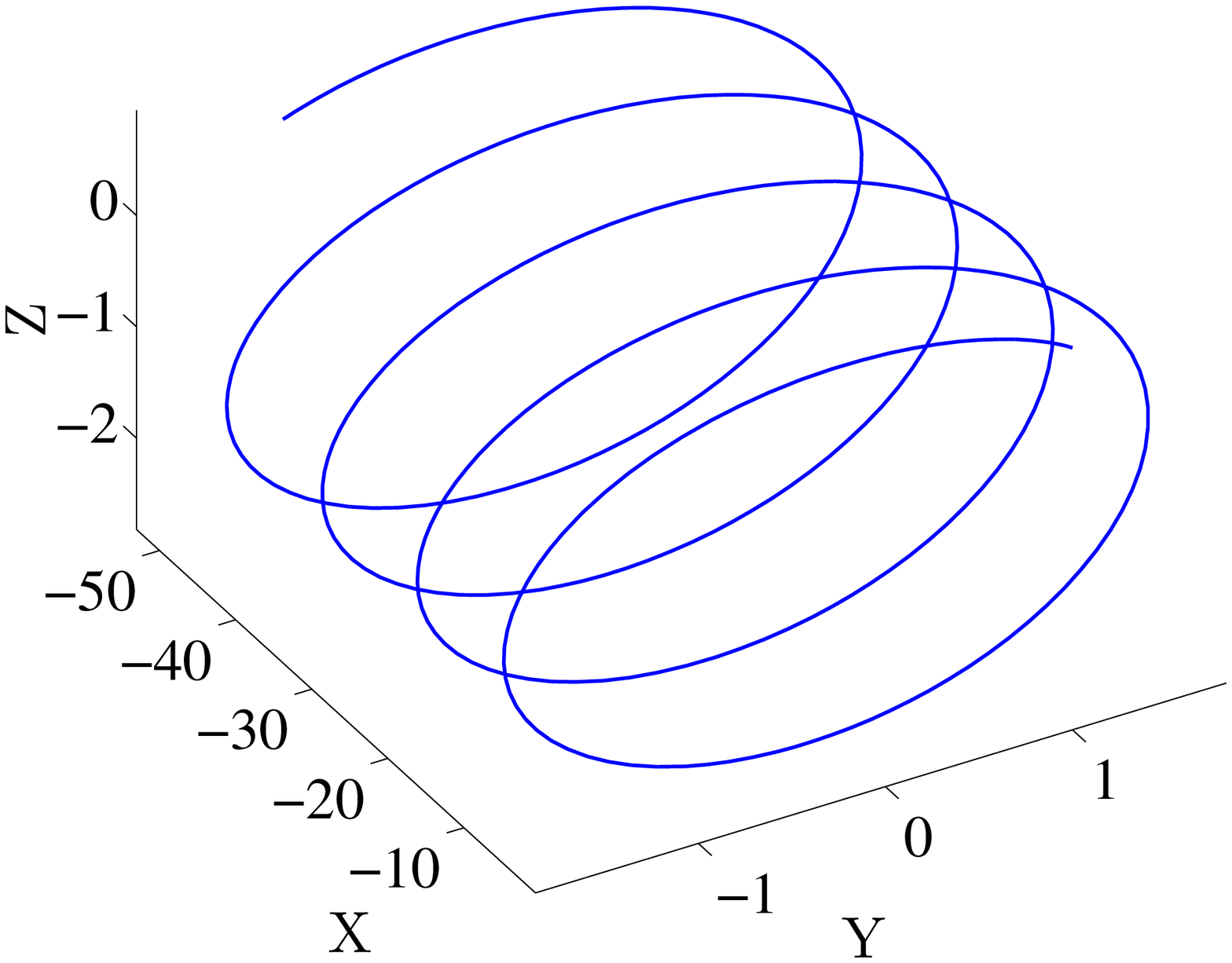}
\includegraphics[width=8cm]{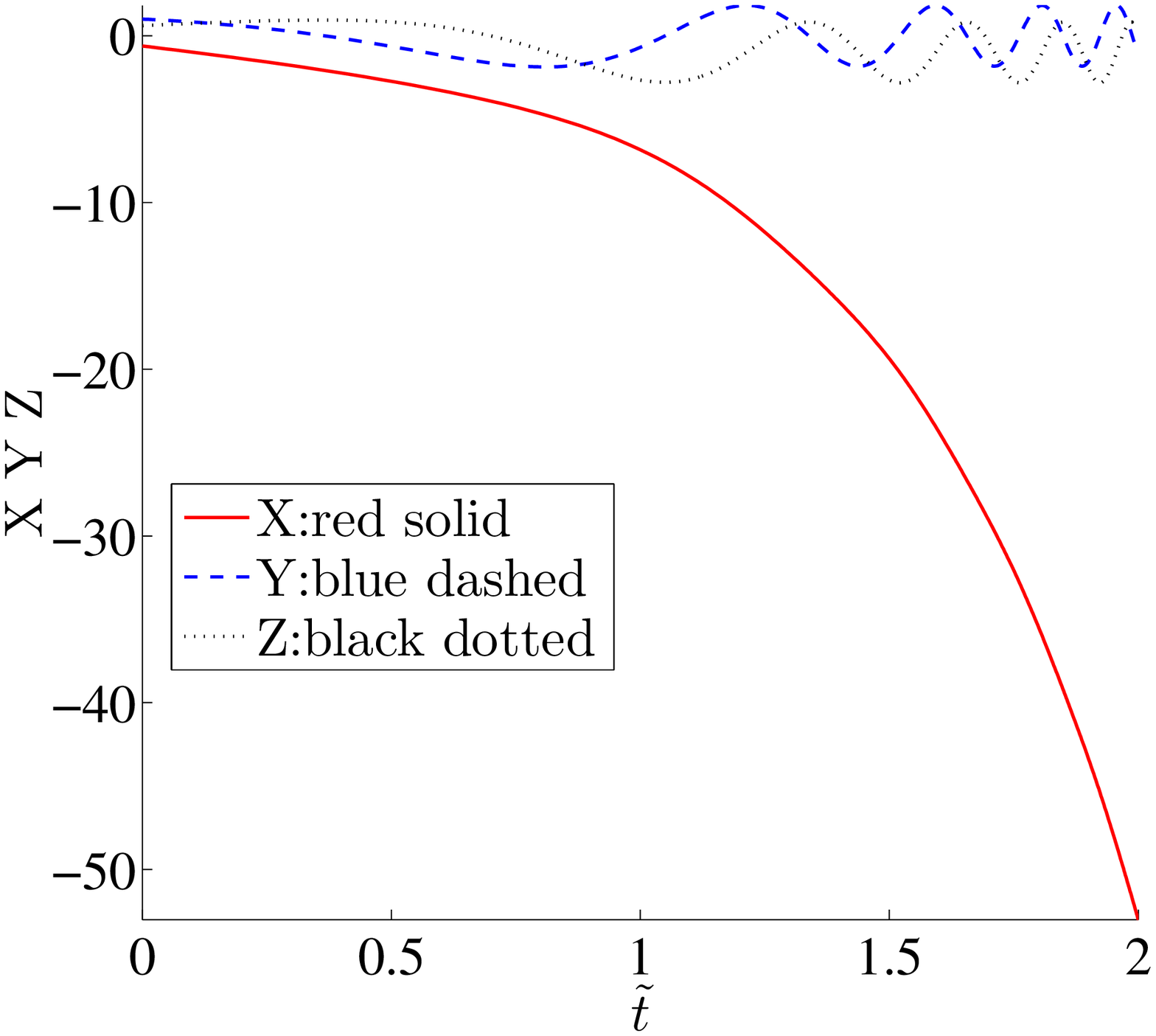} \\
\includegraphics[width=8cm]{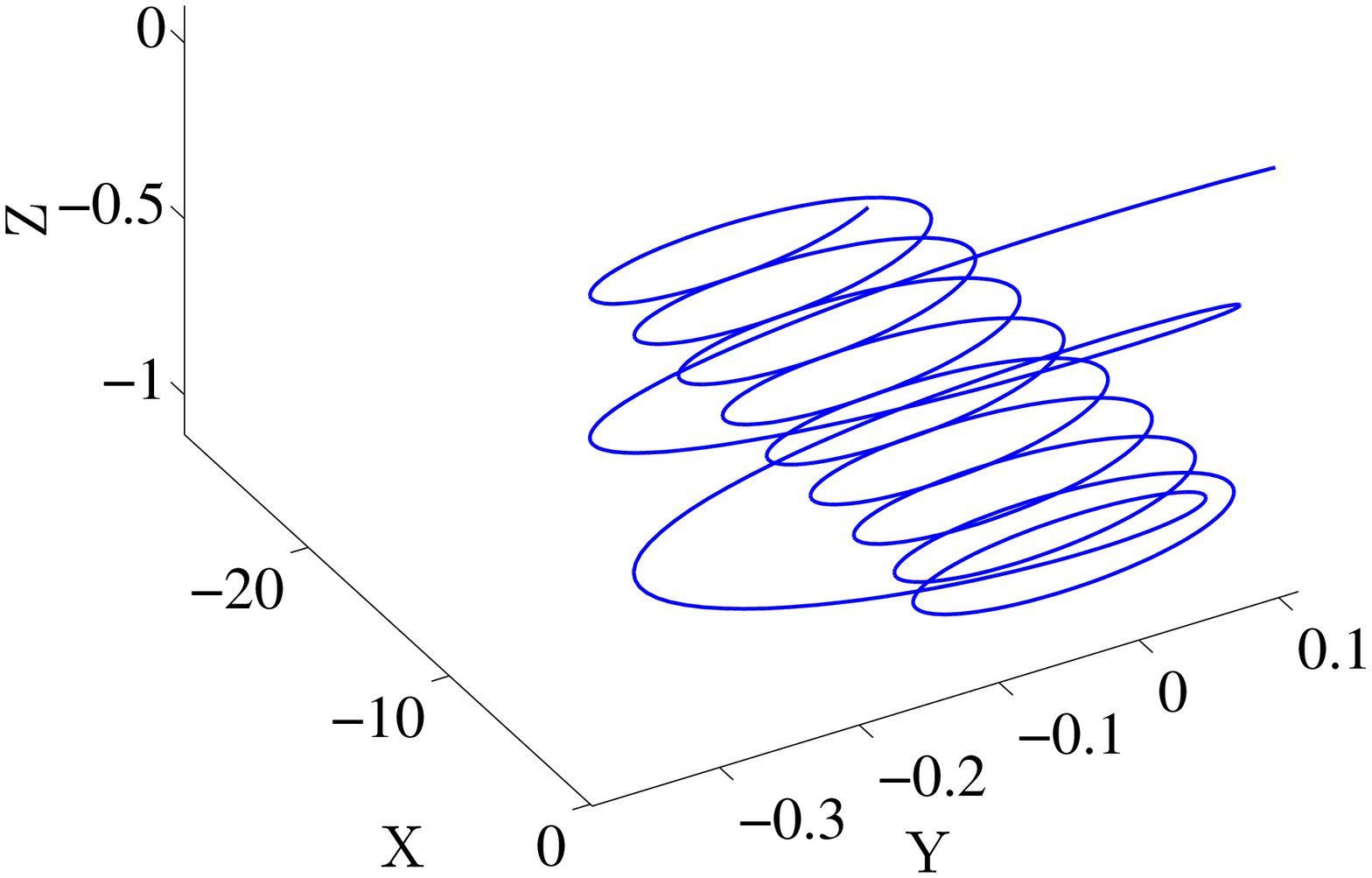}
\includegraphics[width=8cm]{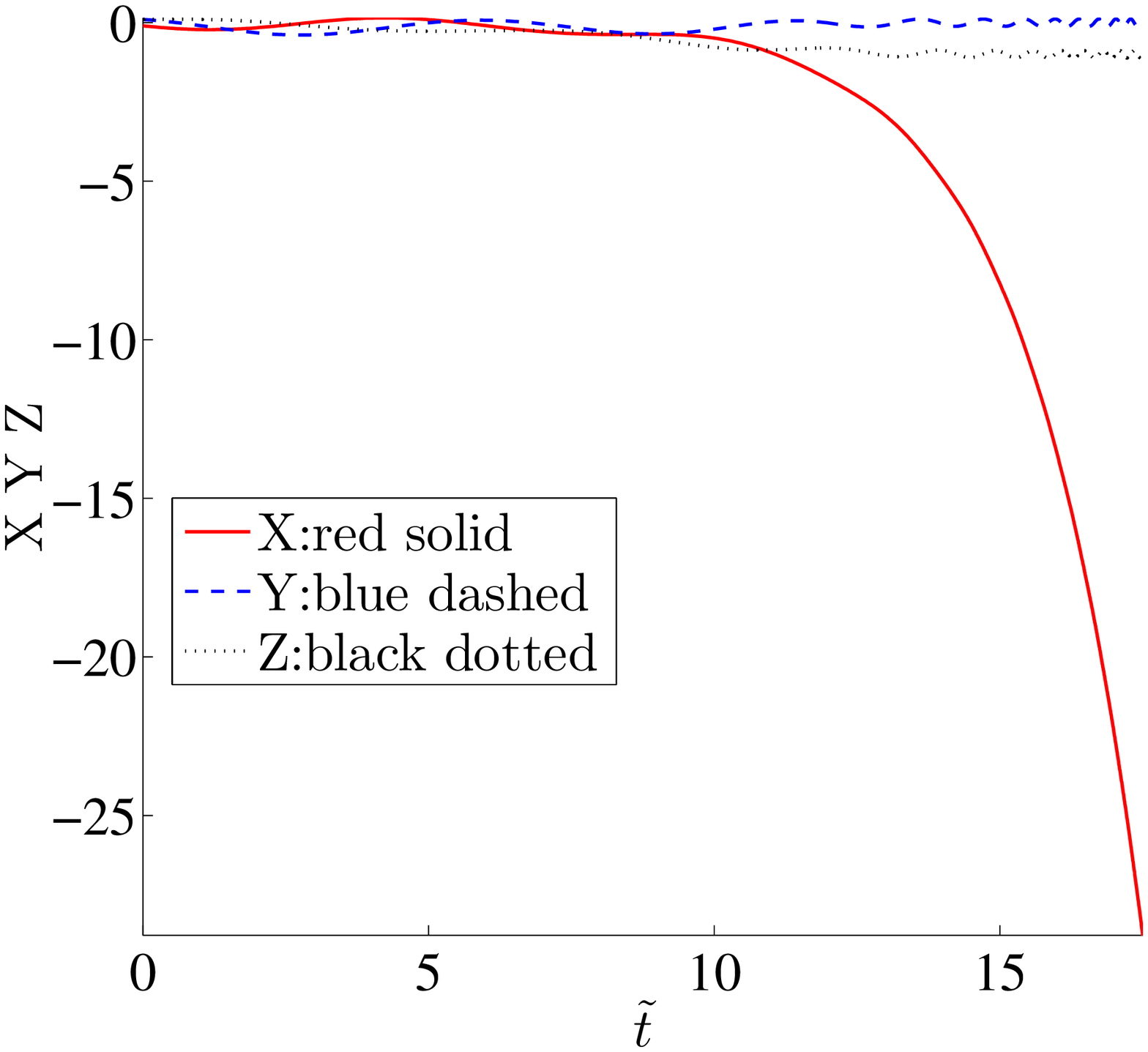}
\end{tabular}
\caption{The top panels are related to the case $b^2>\sigma^2$ with $b=-2$, $\sigma=1$, $P=-1$, starting near the origin $(0,0,0)$ in the Stokes variables and ending up with $X\to-\infty$. The bottom panels
are for the case of
$\sigma^2>b^2$ and $b=-0.5$, $\sigma=1$, $P=-1$. Here the trajectory still escapes to infinity, instead of forming a closed orbit.
The left panels both here and in subsequent plots show the evolution
in the three-dimensional phase space of $X$, $Y$ and $Z$, while the right
panels show the evolution of each of these Stokes variables over time.}
\label{fig2_c}
\end{figure}

By constructing the special function $L=NX-\frac{\sigma}{P}Z$, we can show $\frac{d}{d\tilde{t}}L=b(N^2+X^2)$ so that $L$ is monotonic for nonzero $b$. This confirms that the origin is the only fixed point. Moreover, the system cannot bear any periodic orbits. If it did, integrating $L$ along the orbit would
never give zero, which is a contradiction to the integrated equation. Therefore, even though
this case has exactly the same linearization around the origin as
the previous
case A.2, the dynamics there can be quite different,
as shown in Fig.~\ref{fig2_c}.

\subsection{$a_1=a_2=a$}

We now turn to the case of the active medium, assuming initially,
as is done also in~\cite{barasflach} (and as is most physically
relevant), an equal effect of the active medium on the two waveguides.

\subsubsection{$b_1=b_2=b$, $P_1=P_2=P$}

In this case,  the system in the Stokes variables reads:
\begin{eqnarray}
\frac{dX}{d\tilde{t}}&=&-\sigma Y+ b X,\\
\frac{dY}{d\tilde{t}}&=&b Y+\sigma X+ P X Z,\\
\frac{dZ}{d\tilde{t}}&=&a N +b Z-P X Y.
\label{eqn2d}
\end{eqnarray}
In this case, the
equations remain unchanged if $(x , y)$ is replaced by $(-x, -y)$,
while the evolution of $N$ yields $\frac{d}{d\tilde{t}}(N)=b N +a Z$. So when $b^2>a^2$, $N$ grows to infinity or decays all the way to zero as time $\tilde{t}$ evolves. For $b^2<a^2$, we go back to the original system~(\ref{eqn1}) and consider two invariant manifolds. If $u_1=u_2=u$, the system can be reduced as $\dot{u}=(a+b+i\sigma)u-iP|u|^2u$ and we can show $\frac{d|u|^2}{dt}=2(a+b)|u|^2$. If $u_1=-u_2=u$, the system reads: $\dot{u}=(-a+b-i\sigma)u-iP|u|^2u$. The dynamical equation of $|u|^2$ becomes $\frac{d|u|^2}{dt}=2(-a+b)|u|^2$. Since $|a|>|b|$, one of these two manifolds features exponentially growing solutions while the other only contains exponentially decaying solutions. To be more specific, we can actually give explicit forms for these bounded and unbounded solutions:
\begin{eqnarray}
u_1=u_2=C e^{(a+b)t+i\phi(t)}, ~~~\phi(t)=\sigma t-\frac{P}{2(a+b)}e^{2(a+b)t}+\phi_0
\label{eqn2d1}
\end{eqnarray}
and
\begin{eqnarray}
u_1=-u_2=C e^{(b-a)t+i\phi(t)}, ~~~\phi(t)=-\sigma t-\frac{P}{2(b-a)}e^{2(b-a)t}+\phi_0 ,
\label{eqn2d2}
\end{eqnarray}
where $C$ and $\phi_0$ are constants determined by initial values. The instability of the unbounded solution has been discussed in ~\cite{barasflach2}.

On the other hand, the local analysis yields an interesting cascade
of bifurcations. In particular,
$(0,0,0)$ is always a fixed point;
the eigenvalues of Eq.~(\ref{eqn1}) when
linearizing around the origin are $-a +b -i \sigma$ and $a+ b + i \sigma$.
When $a^2>b^2$, this fixed point becomes unstable, doing so through
a pitchfork bifurcation resulting in the emergence of two new fixed points,
namely
$F^+=(\sqrt{\frac{(a^2-b^2)(b^2+\sigma^2)}{P^2b^2}},\frac{b}{\sigma}\sqrt{\frac{(a^2-b^2)(b^2+\sigma^2)}{P^2b^2}} , -\frac{b^2+\sigma^2}{P\sigma})$ and $F^-=(-\sqrt{\frac{(a^2-b^2)(b^2+\sigma^2)}{P^2b^2}},-\frac{b}{\sigma}\sqrt{\frac{(a^2-b^2)(b^2+\sigma^2)}{P^2b^2}} , -\frac{b^2+\sigma^2}{P\sigma})$.

The characteristic equation for these fixed points $F^+$(or $F^-$) is
$\lambda^3-2b\lambda^2+\frac{(a^2-b^2)\sigma^2}{b^2}\lambda-\frac{(a^2-b^2)(b^2+\sigma^2)}{b}=0$.
When $b^2=\sigma^2$, the Jacobian matrix at $F^+$(or $F^-$) has a real eigenvalue $2b$ and two pure imaginary eigenvalues $\lambda^2=-\frac{\sigma^2(a^2-b^2)}{b^2}=b^2-a^2$. As a result, at this point this system becomes
subject to a Hopf bifurcation giving birth to the existence of
limit cycles.

\begin{figure}[!htbp]
\begin{tabular}{cc}
\includegraphics[width=8cm]{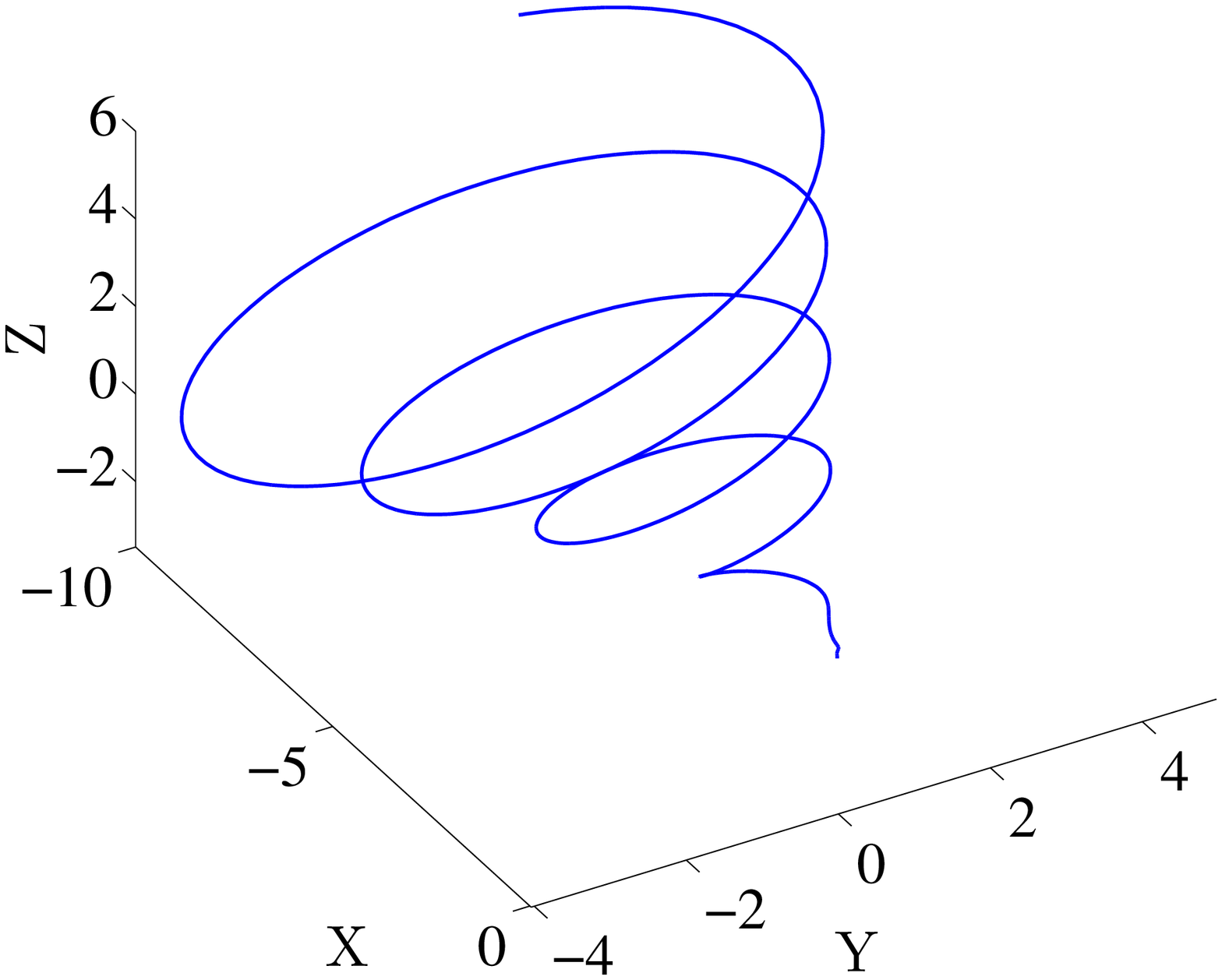}
\includegraphics[width=8cm]{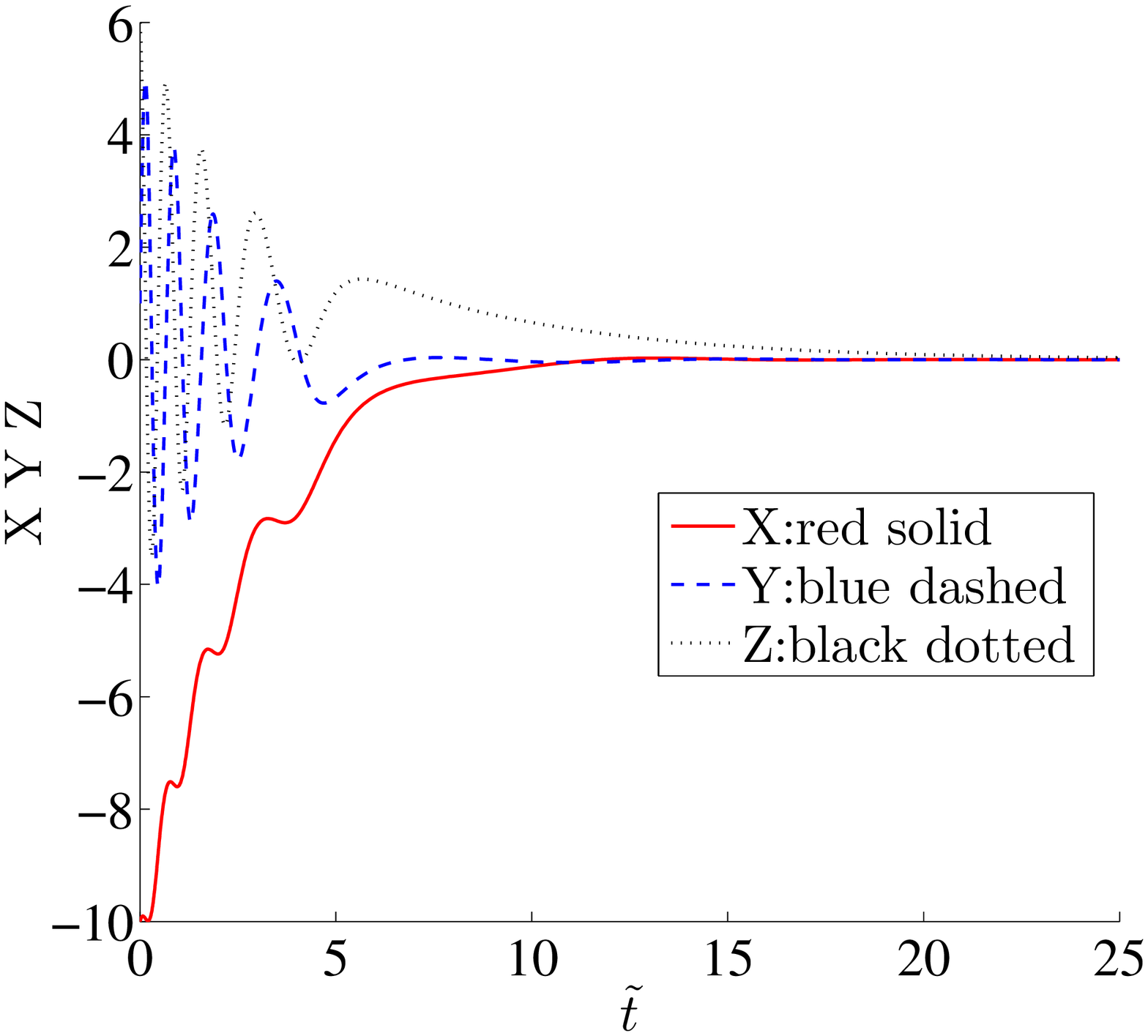} \\
\includegraphics[width=8cm]{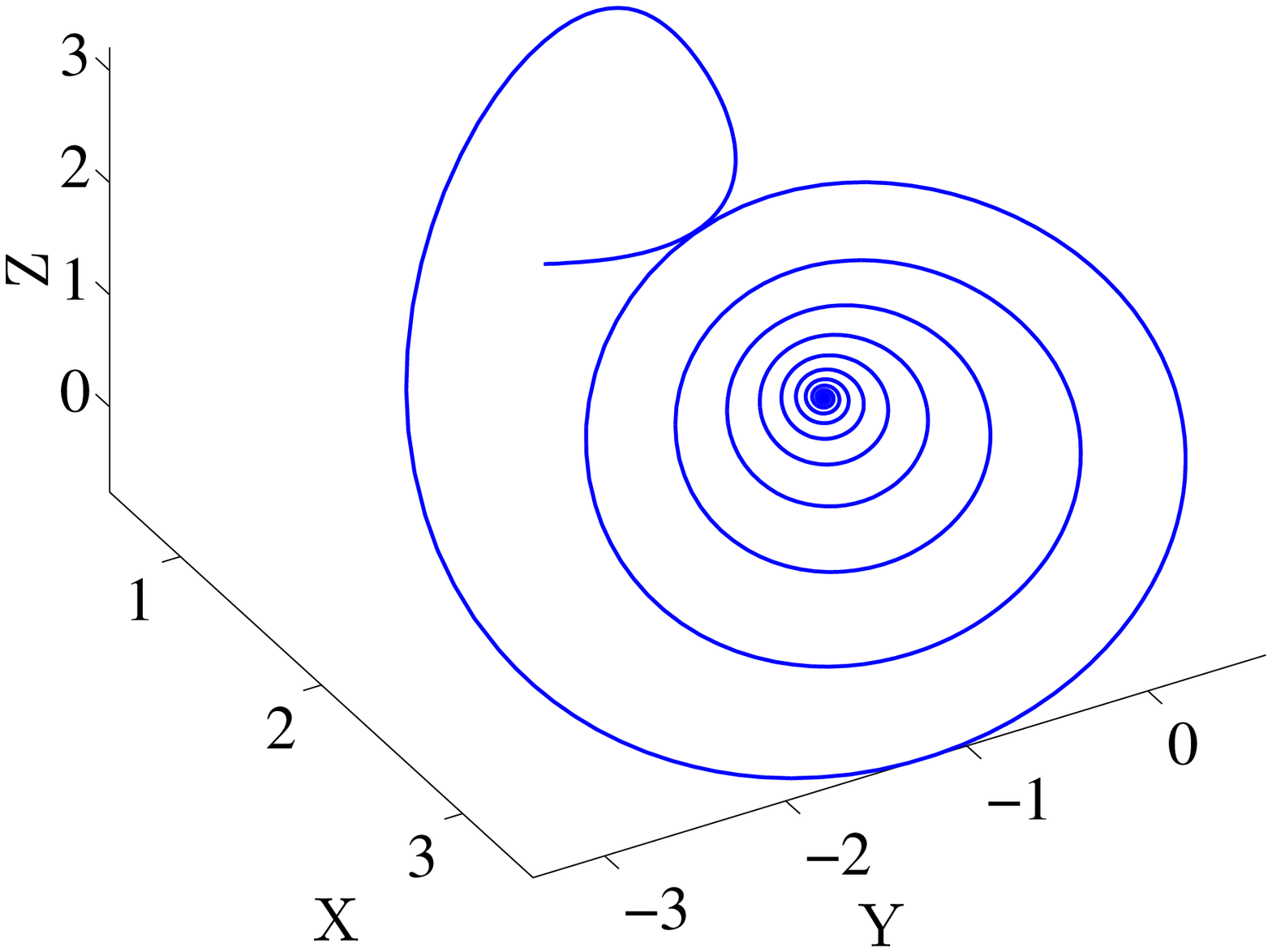}
\includegraphics[width=8cm]{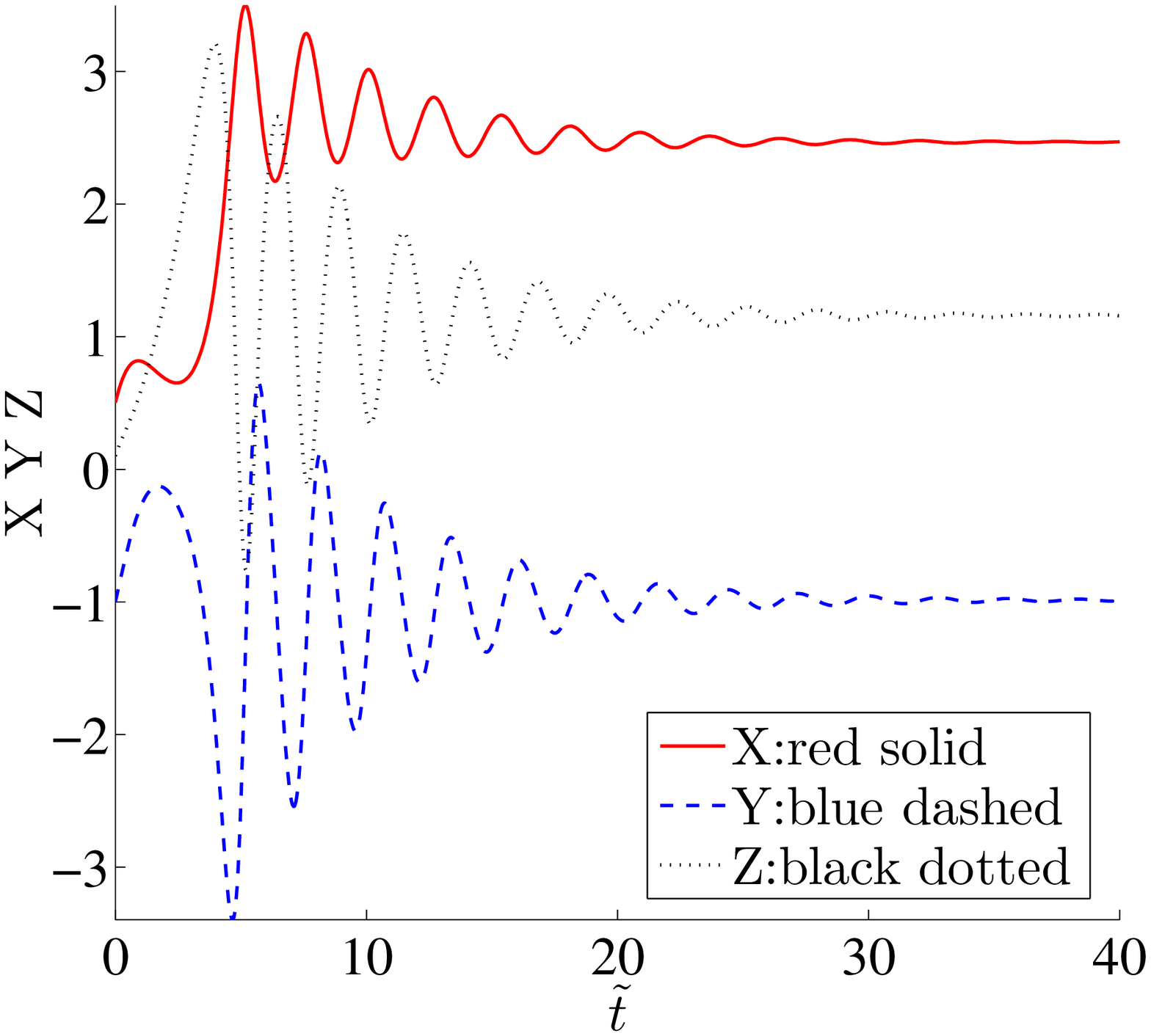}
\end{tabular}
\caption{The top panels show the case (for the active system)
of  $a^2<b^2$ with $a=0.2$, $b=-0.4$, $\sigma=1$, $P=-1$, resulting into
approach to the origin $(0,0,0)$ in the Stokes variables. The bottom panels
are for the supercritical case of
$a^2>b^2$ with $b^2<\sigma^2$ and $a=1$, $b=-0.4$, $\sigma=1$, $P=-1$.
Here the stable steady state $F^+$ is approached (as the origin has
become unstable).}
\label{fig3}
\end{figure}

\begin{figure}[!htbp]
\begin{tabular}{cc}
\includegraphics[width=8cm]{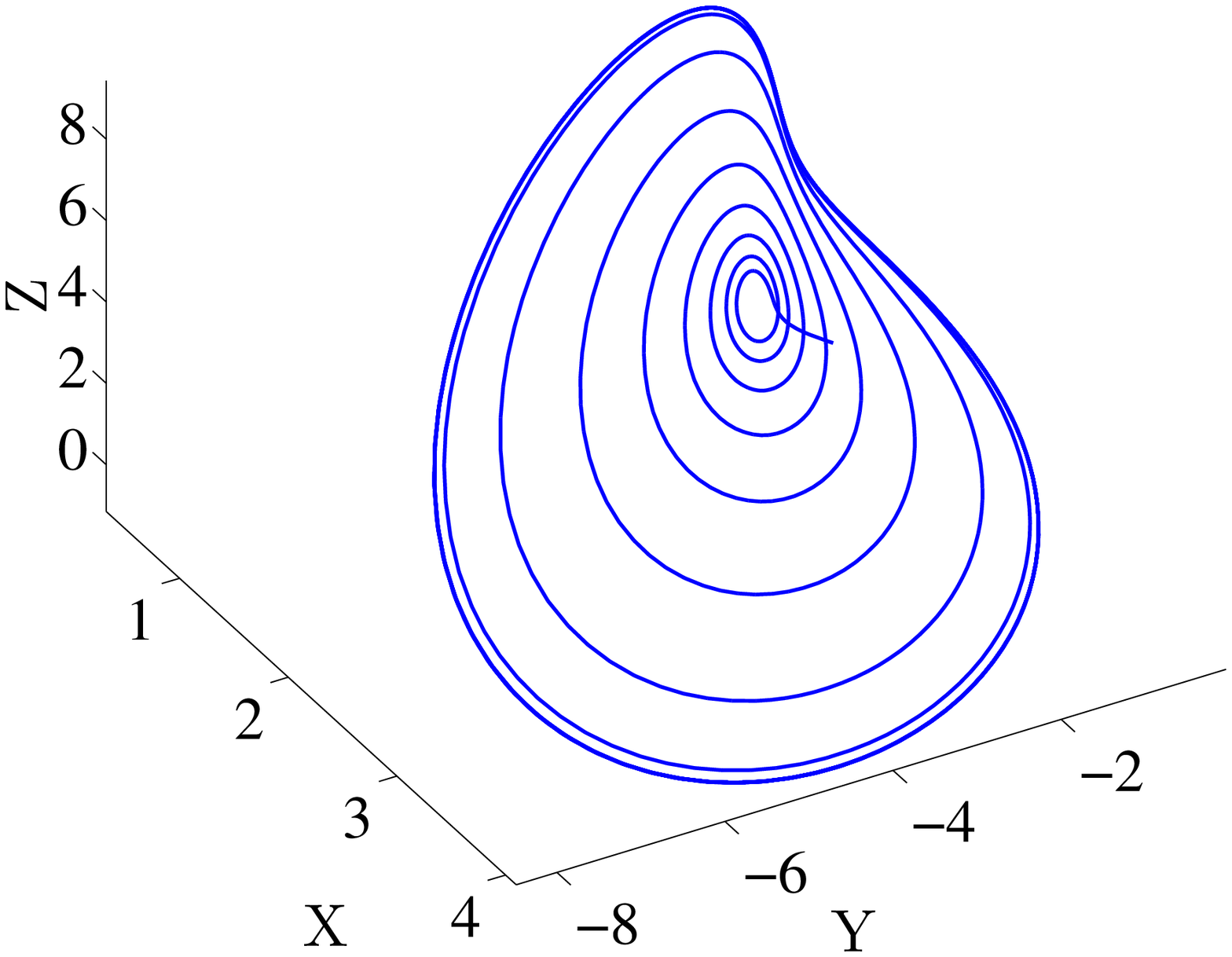}
\includegraphics[width=8cm]{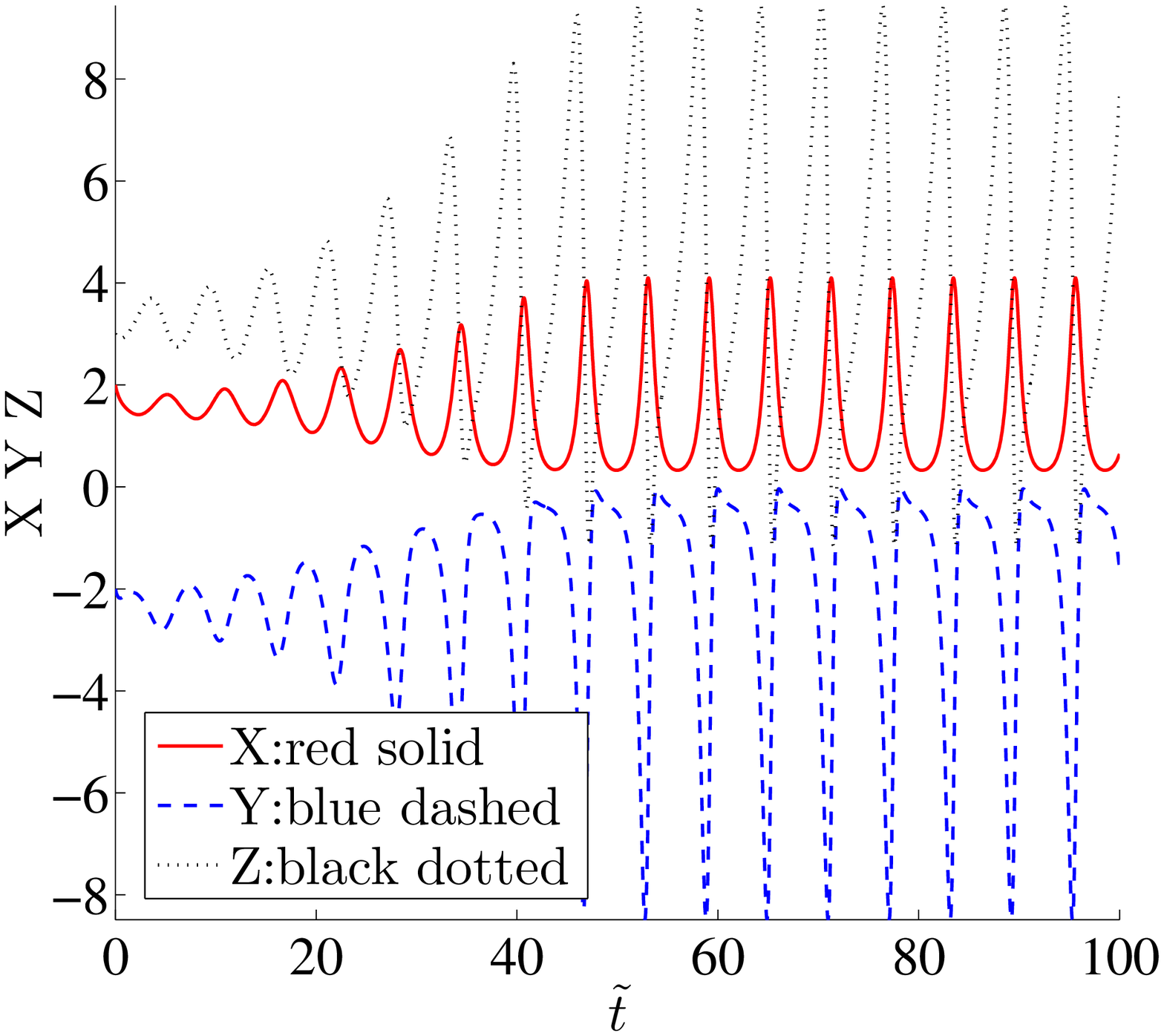} \\
\includegraphics[width=8cm]{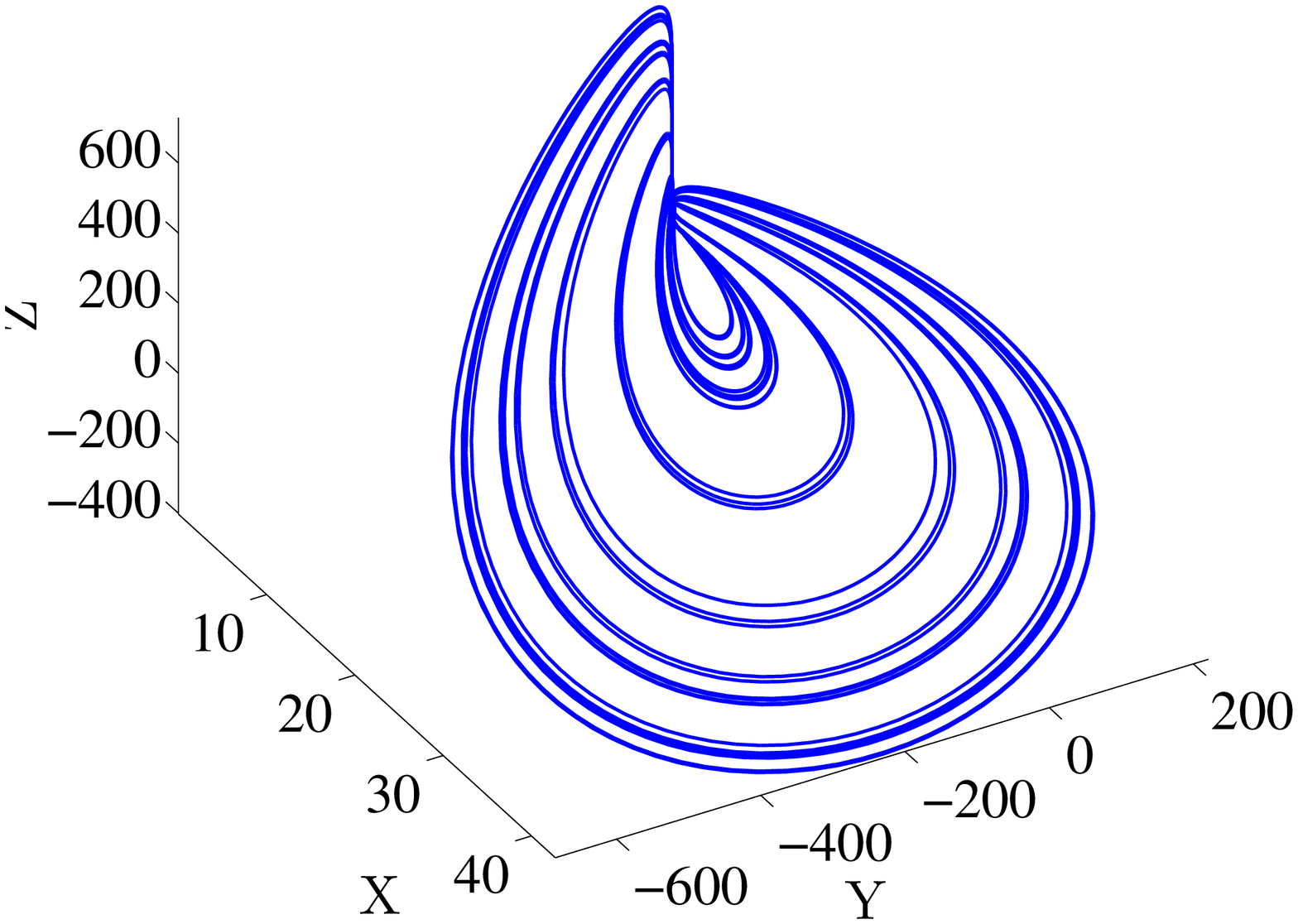}
\includegraphics[width=8cm]{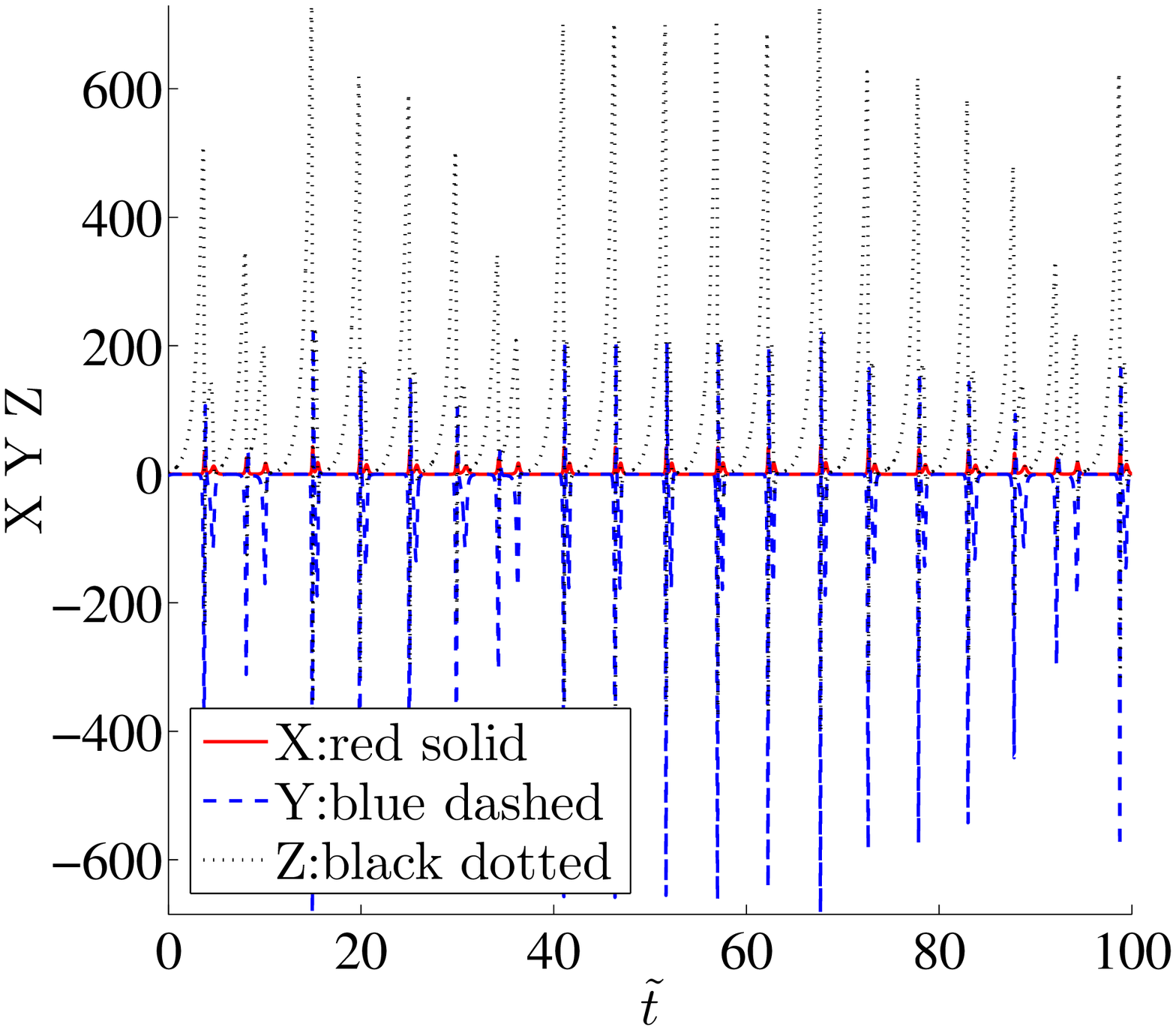}
\end{tabular}
\caption{The top two panels show the case where
$a^2>b^2$ with $b^2>\sigma^2$: $a=2$, $b=-1.5$, $\sigma=1$, $P=-1$.
Here the orbit locks into a limit cycle (as we are past the Hopf
bifurcation point destabilizing $F^{\pm}$). Finally, the bottom
two panels show the case of substantially larger values of $b^2$
with   $a^2>b^2$ and $b^2>\sigma^2$ for $a=9$, $b=-7.6$, $\sigma=1$, $P=-1$.
Here the response of the system is chaotic.}
\label{fig4}
\end{figure}

Subsequent period doublings of the limit cycles yield progressively a
route to chaotic dynamics, as was systematically analyzed for this
case in~\cite{barasflach}. Hence, in this case the dynamics of the
system is very rich. In fact, it transitions
from asymptoting to $(0,0,0)$ to asymptoting
to a stable steady state (through a pitchfork) [these two scenaria
are illustrated in Fig.~\ref{fig3}] and then from asymptoting to a
 periodic orbit eventually to a fully chaotic dynamics [these two
scenaria are shown in Fig.~\ref{fig4}]. These findings
corroborate the analysis of the pioneering study in this generalized
context
of active media of~\cite{barasflach}.


\subsubsection{$b_1=b_2=b$, $P_1=-P_2=P$}

We now turn to another case which constitutes an interesting variant
of the case studied in~\cite{barasflach}. In particular, again
bearing in mind the case of waveguides of different materials with
different nonlinear properties, we consider the case where $P_1=-P_2=P$,
i.e., one waveguide with a focusing and one with a self-defocusing
nonlinearity.

The system in this case is of the form:
\begin{eqnarray}
\frac{dX}{d\tilde{t}}&=&-\sigma Y+ b X,\\
\frac{dY}{d\tilde{t}}&=& b Y+ \sigma X+ P N Z,\\
\frac{dZ}{d\tilde{t}}&=& a N + b Z - P N Y.
\label{eqn2f}
\end{eqnarray}
The stability analysis of the origin in this case is the same
as in case B.1. Moreover, when $(a^2-b^2)>0$, there is also $F=(\frac{\sigma(a^2-b^2)}{Pab}, \frac{(a^2-b^2)}{Pa}, -\frac{b}{a}\sqrt{\frac{(a^2-b^2)(b^2+\sigma^2)}{P^2b^2}})$. The linearization around this fixed point leads to the
polynomial equation (for the eigenvalues):
$\lambda^3-2b\lambda^2+\lambda(\frac{a^2(\sigma^2+b^2)}{b^2})+\frac{(b^2+\sigma^2)(b^2-a^2)}{b}=0$. If the equation has three real roots, they are always negative~($b<0$) or positive~($b>0$). If two of the roots are complex conjugates and one is real, it can be easily checked that the real root always has the same sign as $b$. By expressing $(\sigma^2+b^2)$ as a function of three roots and using the fact it is positive, we find here the real part of
the complex roots also shares the sign of $b$. Thus we find that $a^2>b^2$ changes the origin to a saddle and admits the emergence of a stable~(if $b<0$) or unstable~(if $b>0$) fixed point in $F$.

\begin{figure}[!htbp]
\begin{tabular}{cc}
\includegraphics[width=8cm]{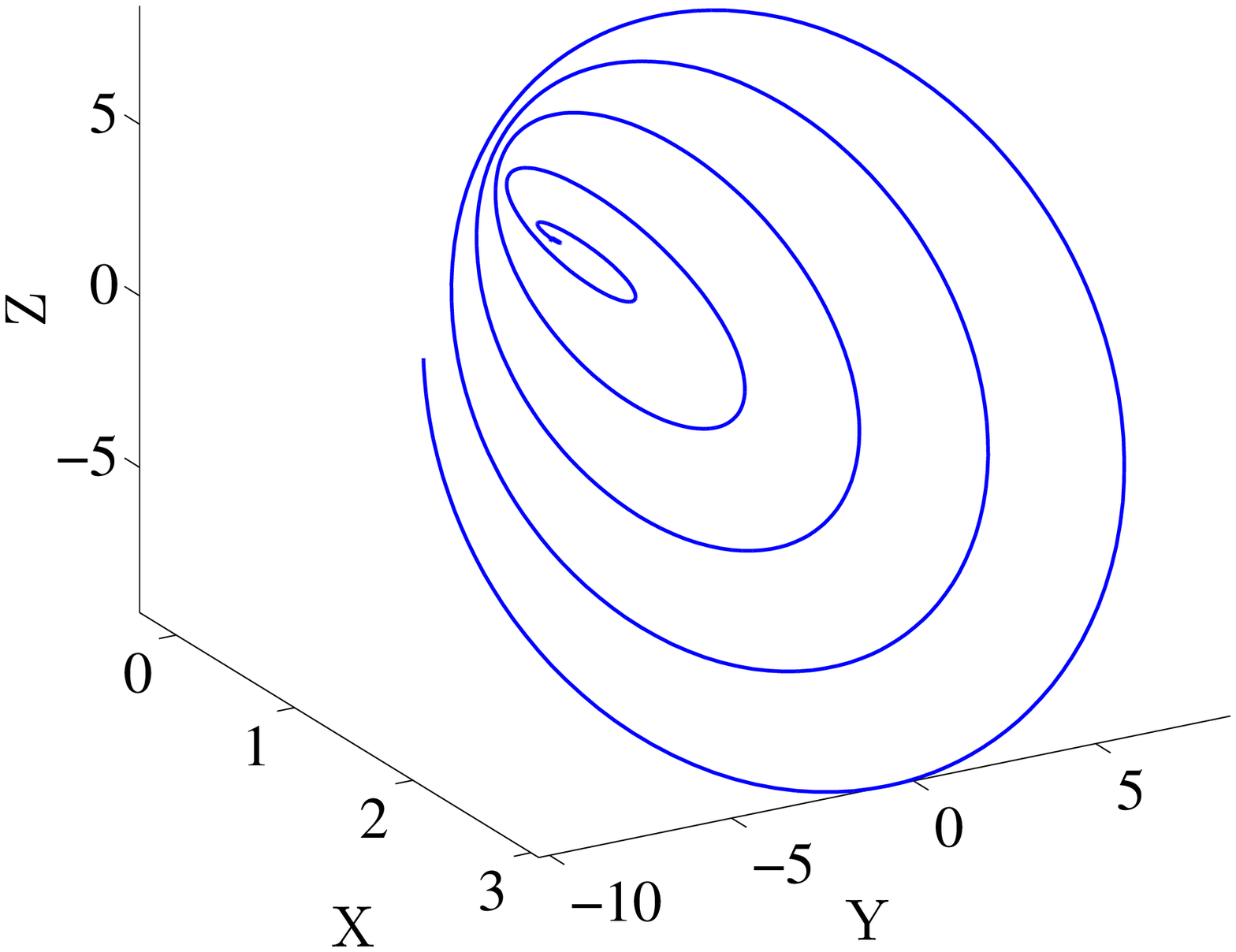}
\includegraphics[width=8cm]{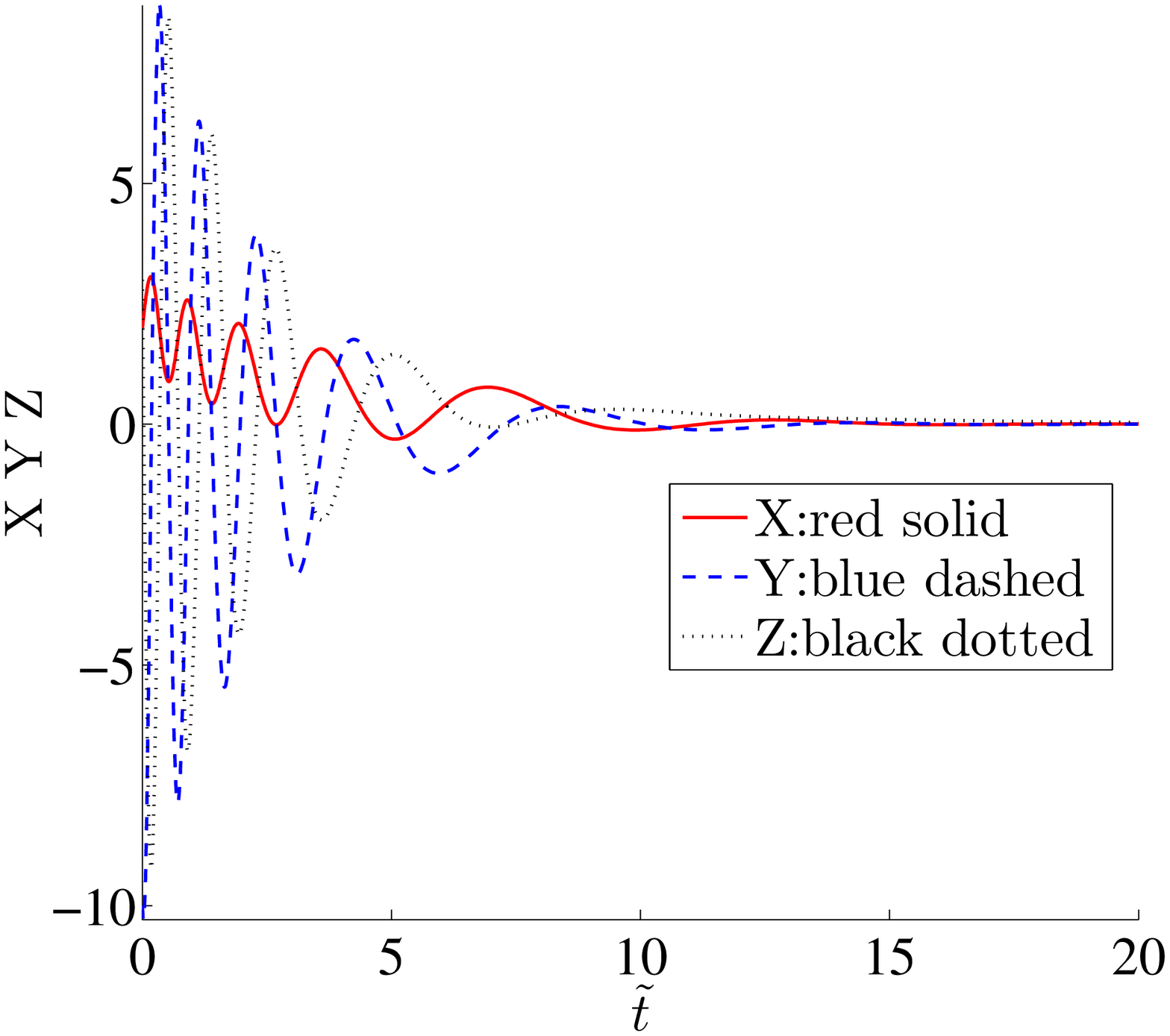} \\
\includegraphics[width=8cm]{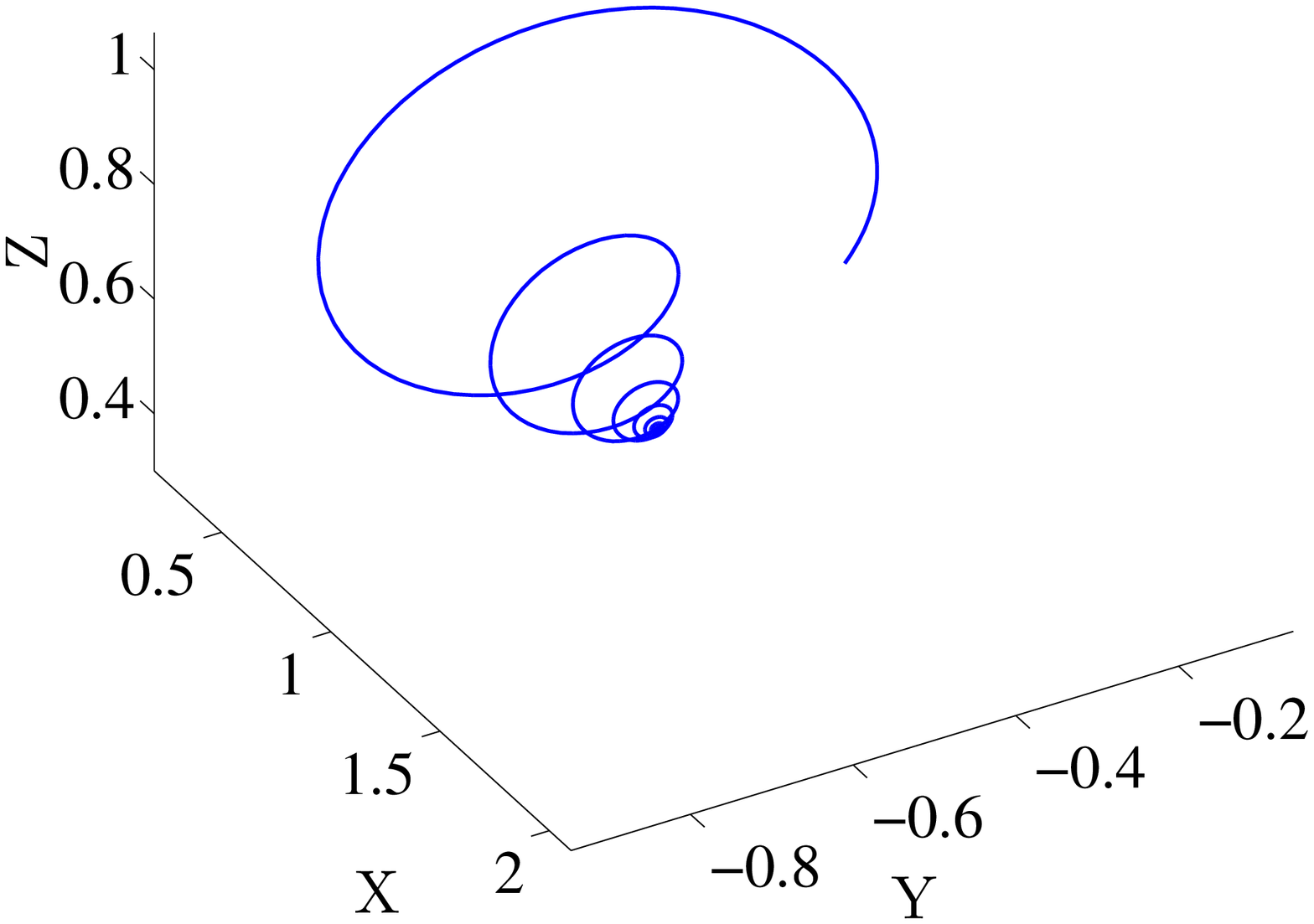}
\includegraphics[width=8cm]{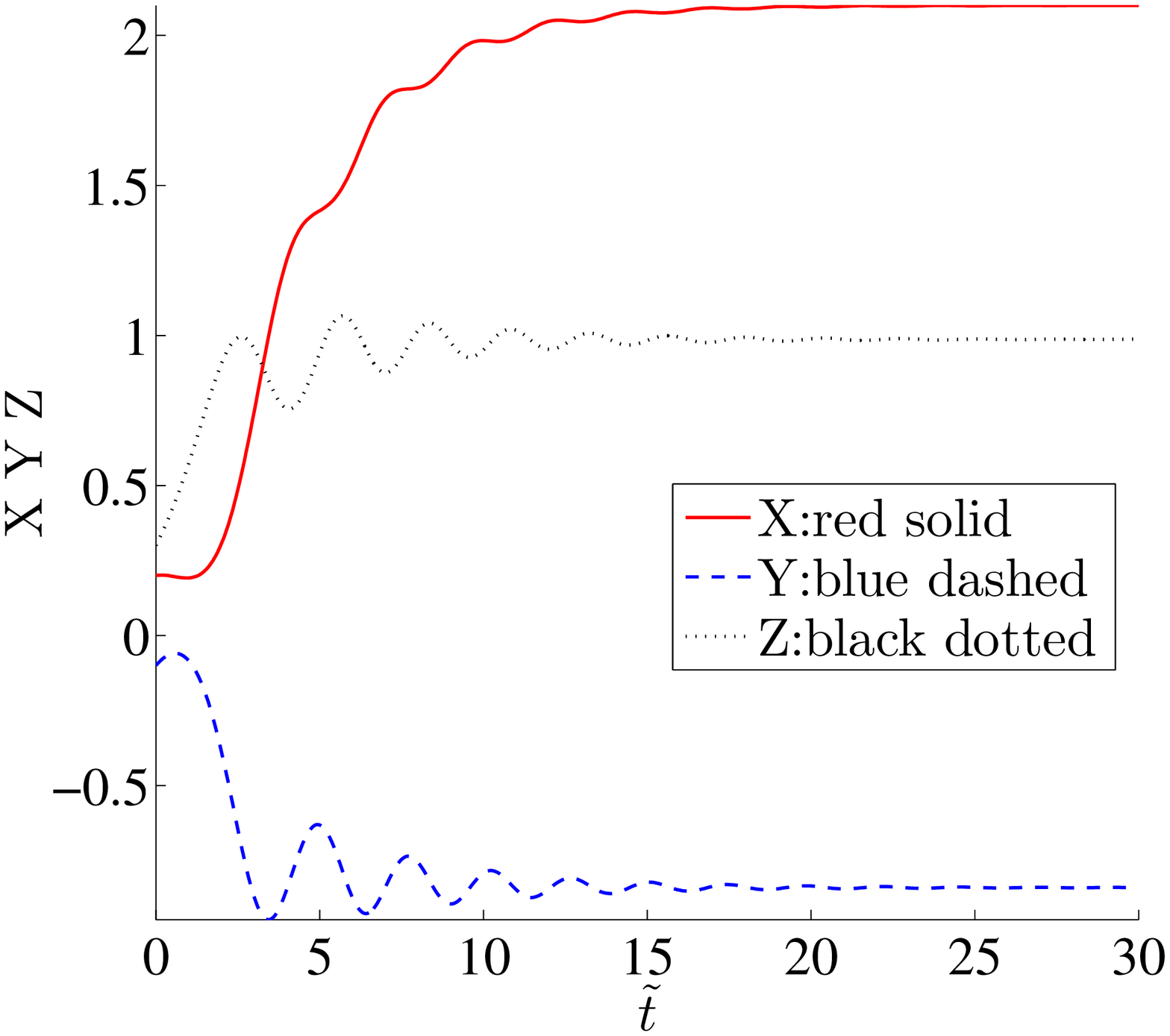}
\end{tabular}
\caption{Top panels: case of $a^2 < b^2$ with $a=0.5$, $b=-1.5$, $\sigma=1$, $P=1$
leading to decay towards the origin. Bottom panels: case of $a^2>b^2$ with
$a=0.5$, $b=-0.2$, $\sigma=1$, $P=1$. Here the origin is unstable
and the evolution leads to the stable nontrivial fixed point $F$.}
\label{fig6}
\end{figure}

These stability conclusions are mirrored in the dynamics of Fig.~\ref{fig6}.
For $a^2<b^2$ (top panels),
we observe that the evolution is driven to the origin,
while for $a^2>b^2$ (bottom panels),
the dynamics leads to the stable steady state $F$
(for our case of $b<0$). Note that if we change $b$ from negative to positive, the evolution will do something opposite to the above figures, spiraling out from the fixed point
(such as the origin in the top panels or $F$ in the bottom panels)
towards infinity.






\subsubsection{$b_1=-b_2=b$, $P_1=P_2=P$}

We now turn to a case slightly different from that of~\cite{barasflach},
yet still physically realizable. In particular, the intrinsic
gain-loss pattern of the waveguides involves gain in one, and loss
in the other (as it does in the $\mathcal{P T}$-symmetric setting),
yet the active medium (and the nonlinearity) uniformly affect both.

The system in this case reads:
\begin{eqnarray}
\frac{dX}{d\tilde{t}}&=&-\sigma Y+ b N,\\
\frac{dY}{d\tilde{t}}&=& \sigma X+ P X Z,\\
\frac{dZ}{d\tilde{t}}&=& a N - P X Y.
\label{eqn2e}
\end{eqnarray}
In this case, it is not straightforward to infer dynamical features
from the evolution of $N^2$ according to
$\frac{d}{d\tilde{t}}(N^2)=2b N X + 2a N Z$.
Hence, we turn to the stability of fixed points.
$(0,0,0)$ is always a fixed point and its eigenvalues in this
case acquire a complicated form in Eq.~(\ref{eqn1}) being
given by $\pm \sqrt{a^2 + b^2 - \sigma^2+2 i a \sigma}$, implying the origin is a fixed point of saddle type.

On the other hand,
when $b^2>\sigma^2$, $|bN|>|\sigma Y|$ so that $X$ is always increasing or
decreasing. When $\sigma^2>b^2$, there is an additional fixed point
of the form $F=(\frac{a\sigma}{b P}, {\rm sign}(b\sigma)\sqrt{\frac{\sigma^2(a^2+b^2)}{P^2(\sigma^2-b^2)}} , -\frac{\sigma}{P})$.
The characteristic equation of the Jacobian matrix at $F$
yields
 \begin{eqnarray}
\lambda^3-\lambda\left(\frac{b^4+a^2b^2-a^2\sigma^2}{b^2}\right)-\frac{a\sigma}{b^2P}\sqrt{P^2b^2(\sigma^2-b^2)(a^2+b^2)}=0.
\label{stabf}
\end{eqnarray}
Since the coefficient of second order term is zero, we know immediately that
at least one of the eigenvalues should have a positive real part.
Hence we conclude that the trajectory near $F$ is unbounded in one or two directions given by the corresponding
eigenvectors and this fixed point is never stable.

\begin{figure}[!htbp]
\begin{tabular}{cc}
\includegraphics[width=8cm]{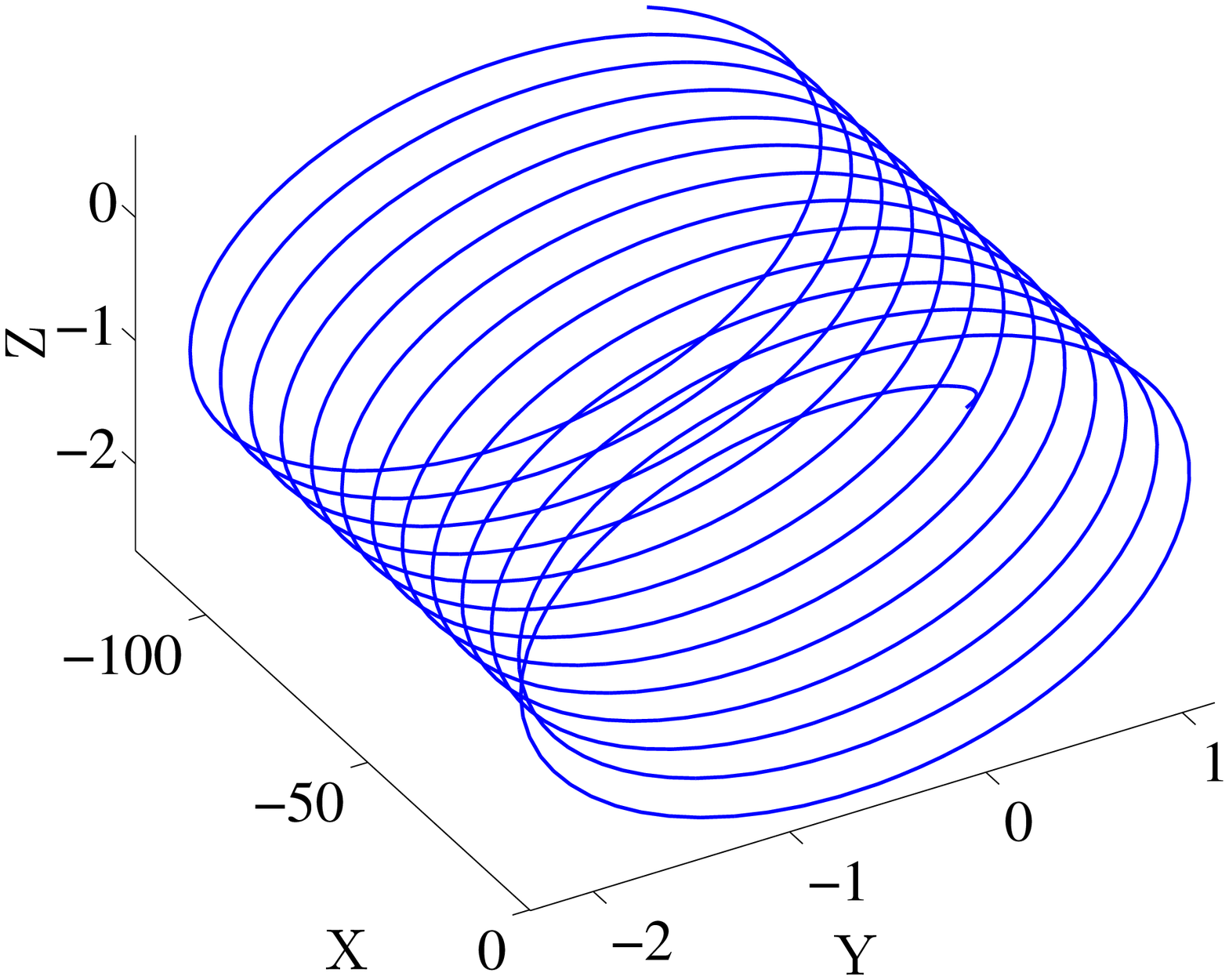}
\includegraphics[width=8cm]{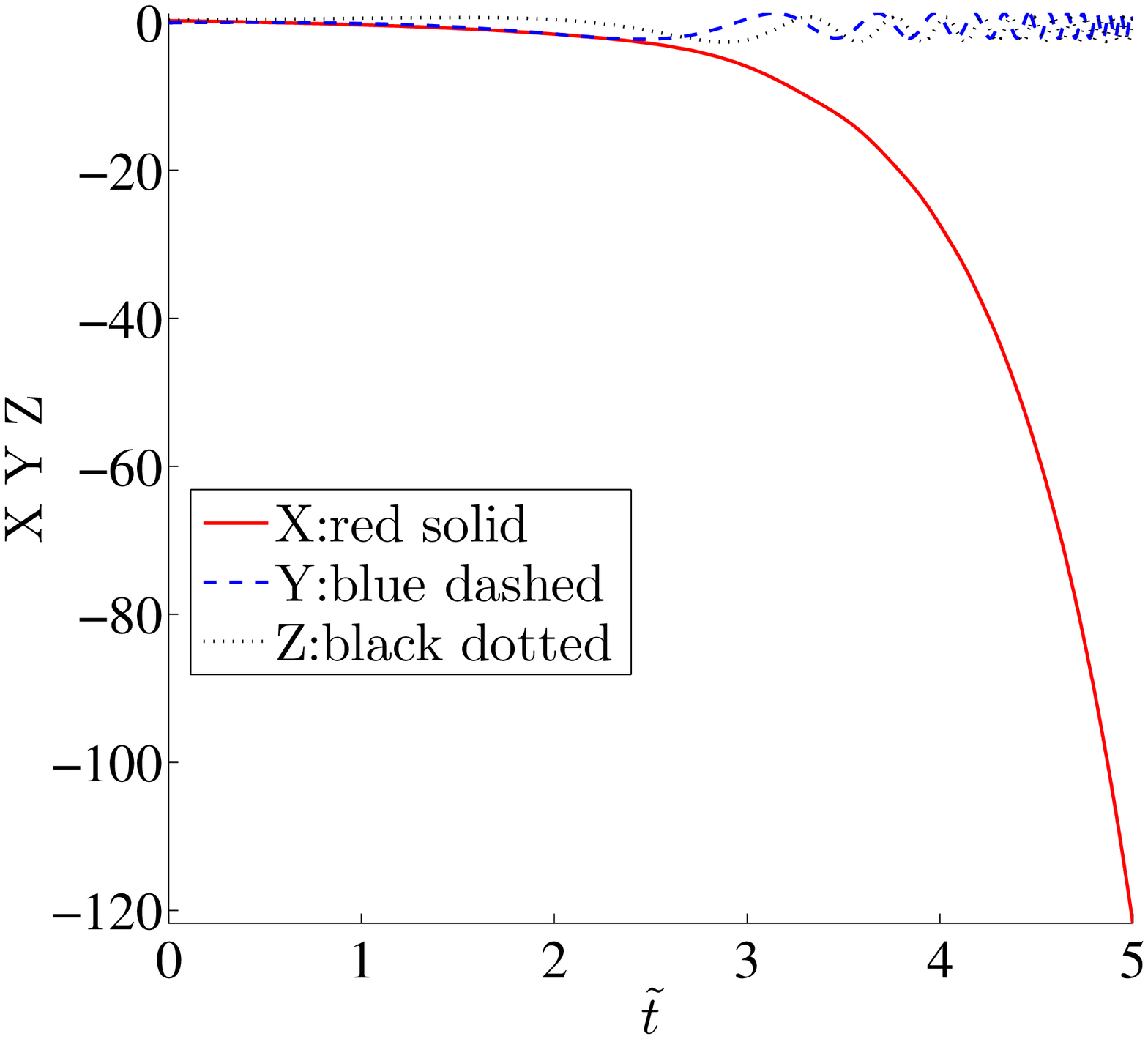} \\
\includegraphics[width=8cm]{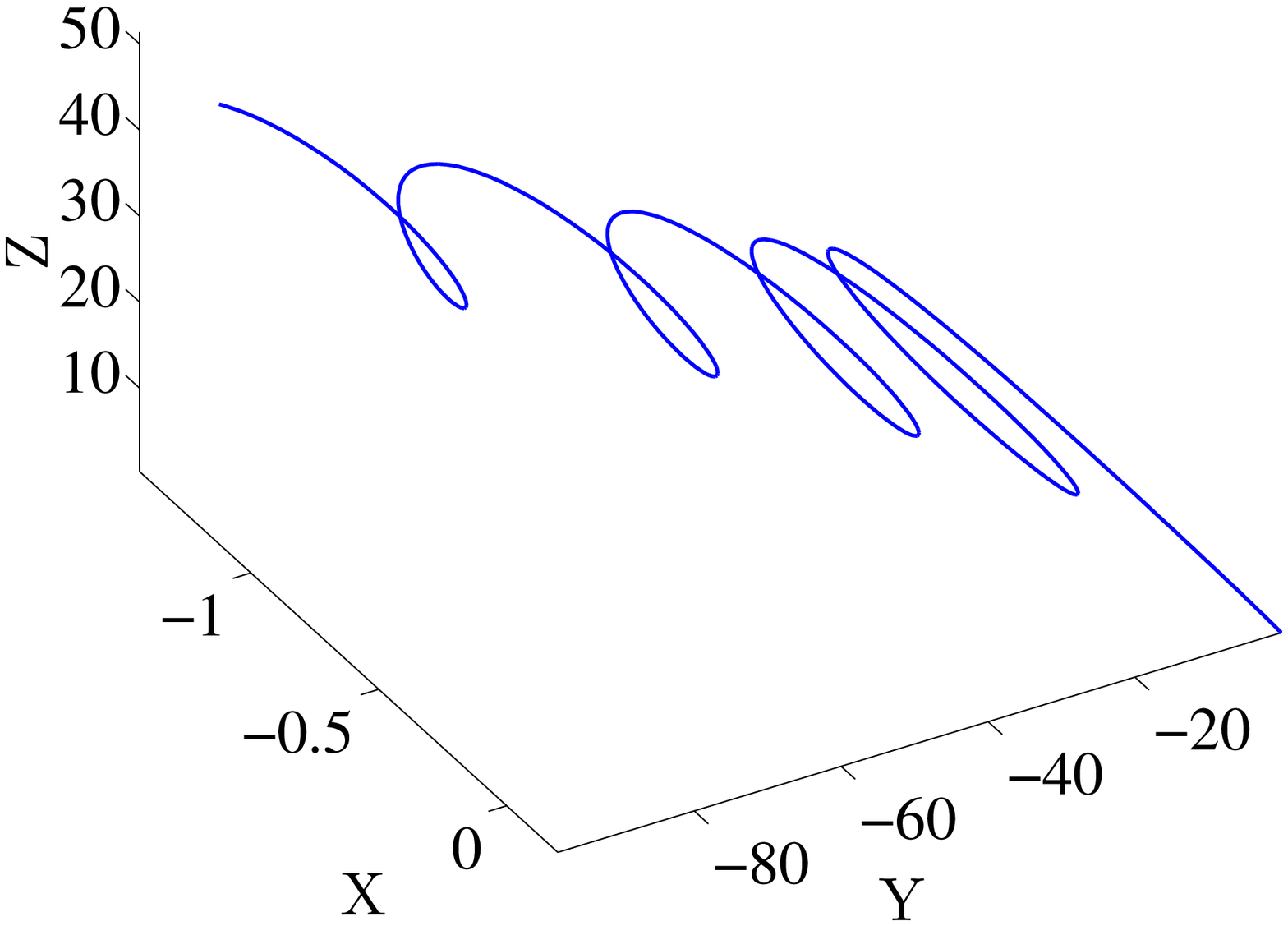}
\includegraphics[width=8cm]{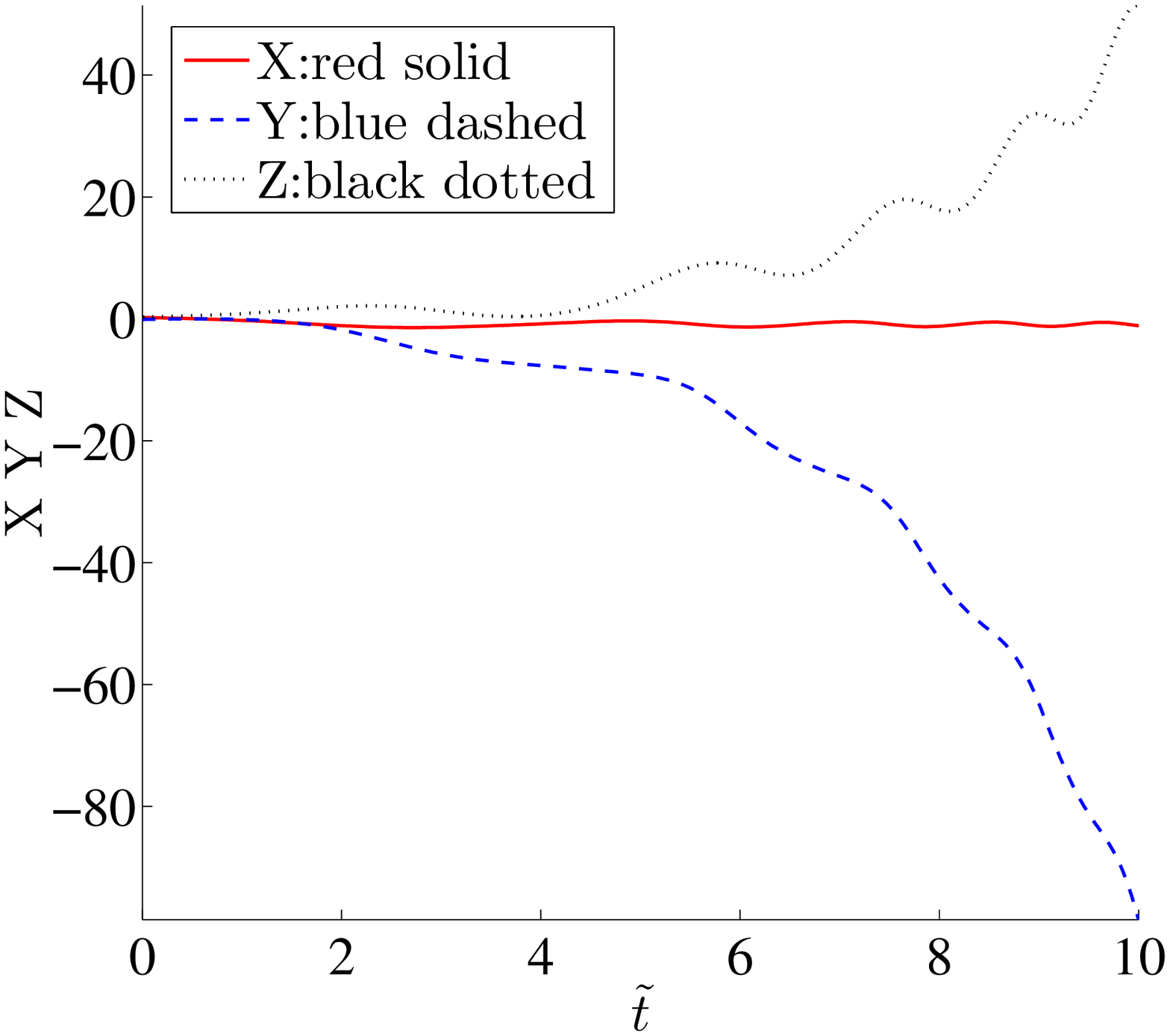}
\end{tabular}
\caption{The top panel contains the case of $\sigma^2<b^2$
with $a=0.5$, $b=-1.5$, $\sigma=1$, $P=1$. The origin is the sole (yet
unstable) fixed point here with the dynamics leading $X \to -\infty$,
while  $Y$ and $Z$ will circle around $Y=-0.5, Z=-1$. On the other hand,
the bottom panels are for the case with $\sigma^2 > b^2$ with
$a=1$, $b=-0.9$, $\sigma=1$, $P=1$. Here $Y,Z$ tend to $\pm \infty$,
while $X$ stays bounded. The latter implies that the amplitudes of both
waveguides tend to $\infty$.}
\label{fig5}
\end{figure}

Despite the existence of such an additional fixed point,
we thus find in this case that the generic
instability of $F$, as well as that of the origin lead the dynamics
to feature an evolution of $|X|\to\infty$ or of $|Y|,|Z|\to\infty$.
This is clearly demonstrated in Fig.~\ref{fig5}, showcasing
the relevant dynamics both for the case of $\sigma^2 < b^2$ (top panels),
as well as in that of $\sigma^2 > b^2$ (bottom panels).

%

\subsubsection{$b_1=-b_2=b$, $P_1=-P_2=P$}

Finally, we consider the case where features of cases B.2 and
B.3 are combined, namely while the active medium is present, the
waveguides are made of different materials (featuring one with focusing
and the other with self-defocusing nonlinearity), while the intrinsic gain/loss
structure supports one waveguide with gain and features loss in the other.

The system in this case reads:
\begin{eqnarray}
\frac{dX}{d\tilde{t}}&=&-\sigma Y+ b N,\\
\frac{dY}{d\tilde{t}}&=& \sigma X+ P N Z,\\
\frac{dZ}{d\tilde{t}}&=& a N - P N Y.
\label{eqn2g}
\end{eqnarray}
The stability analysis of the origin is similar to the case B.3.

In addition, when $\sigma^2>b^2$ and $\frac{\sigma a}{P b}>0$, there are also $F^+=(\frac{a^2}{Pb}\sqrt{\frac{(\sigma^2-b^2)}{(a^2+b^2)}},\frac{a}{P} , -\frac{a}{P}\sqrt{\frac{(\sigma^2-b^2)}{(a^2+b^2)}})$ and\\  $F^-=(-\frac{a^2}{Pb}\sqrt{\frac{(\sigma^2-b^2)}{(a^2+b^2)}},\frac{a}{P} , \frac{a}{P}\sqrt{\frac{(\sigma^2-b^2)}{(a^2+b^2)}})$. The characteristic equation of Jacobian matrix at $F^+$(or $F^-$) is $\lambda^3+\lambda\frac{a^2\sigma^2+b^2\sigma^2-b^4}{b^2}+\mp\frac{a\sigma\sqrt{(\sigma^2-b^2)(a^2+b^2)}}{b}=0$. By similar analysis as in case B.3, we know at either $F^+$ or $F^-$ one eigenvalue has positive real part and the other two come with negative real parts, while the real parts of eigenvalues at the other fixed point have exactly opposite signs, namely, two positive and one negative.

\begin{figure}[!htbp]
\begin{tabular}{cc}
\includegraphics[width=8cm]{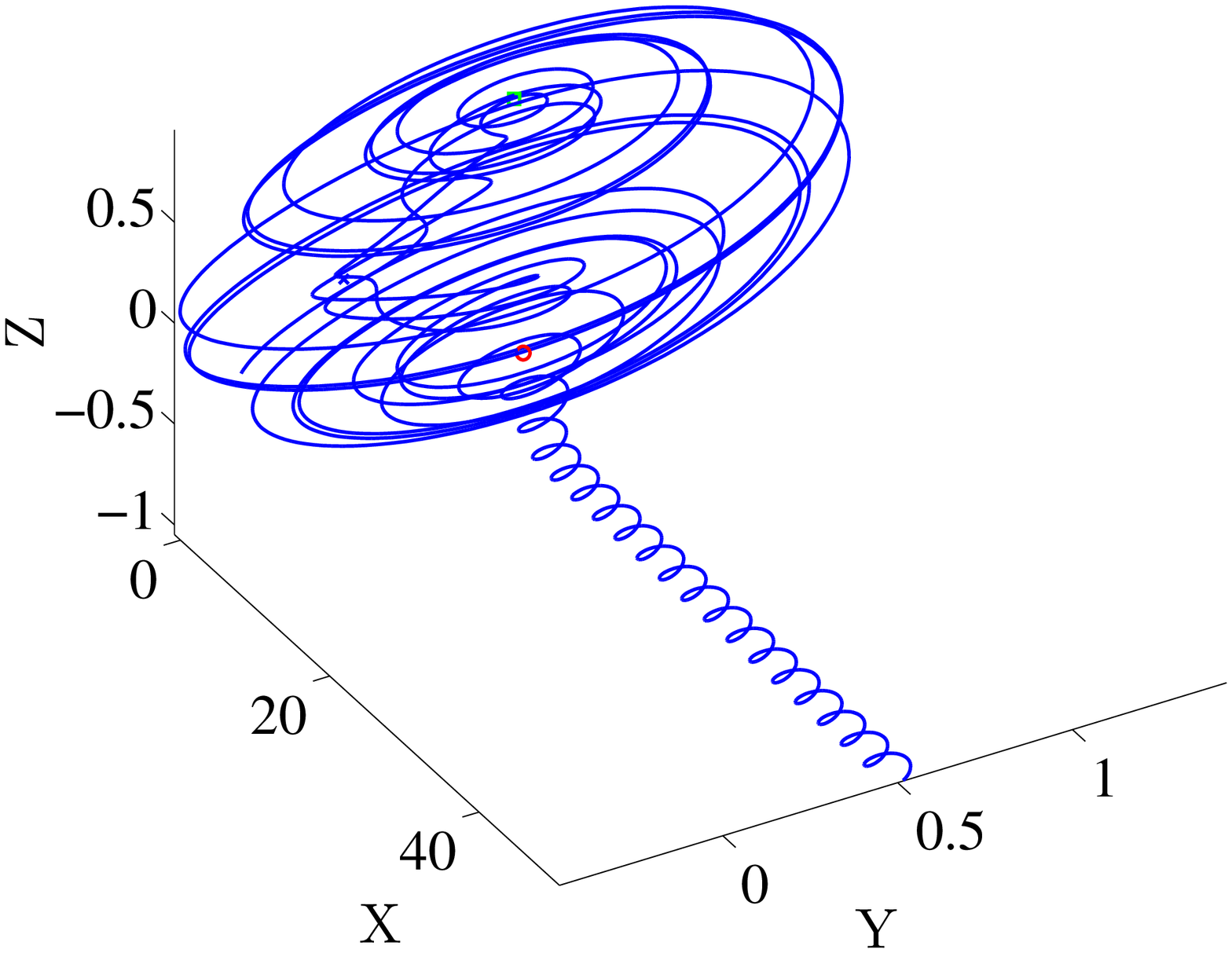}
\includegraphics[width=8cm]{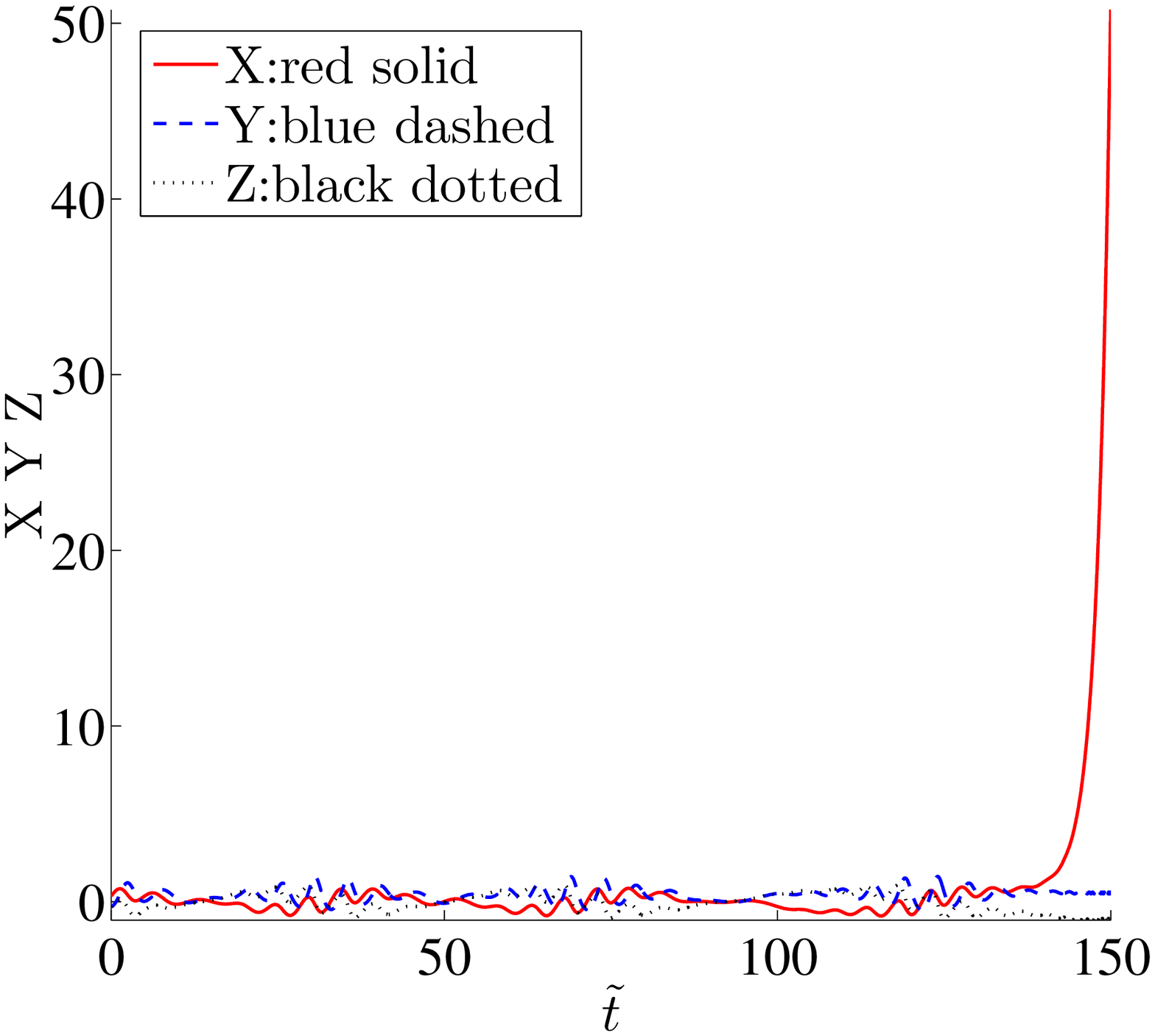}\\
\includegraphics[width=8cm]{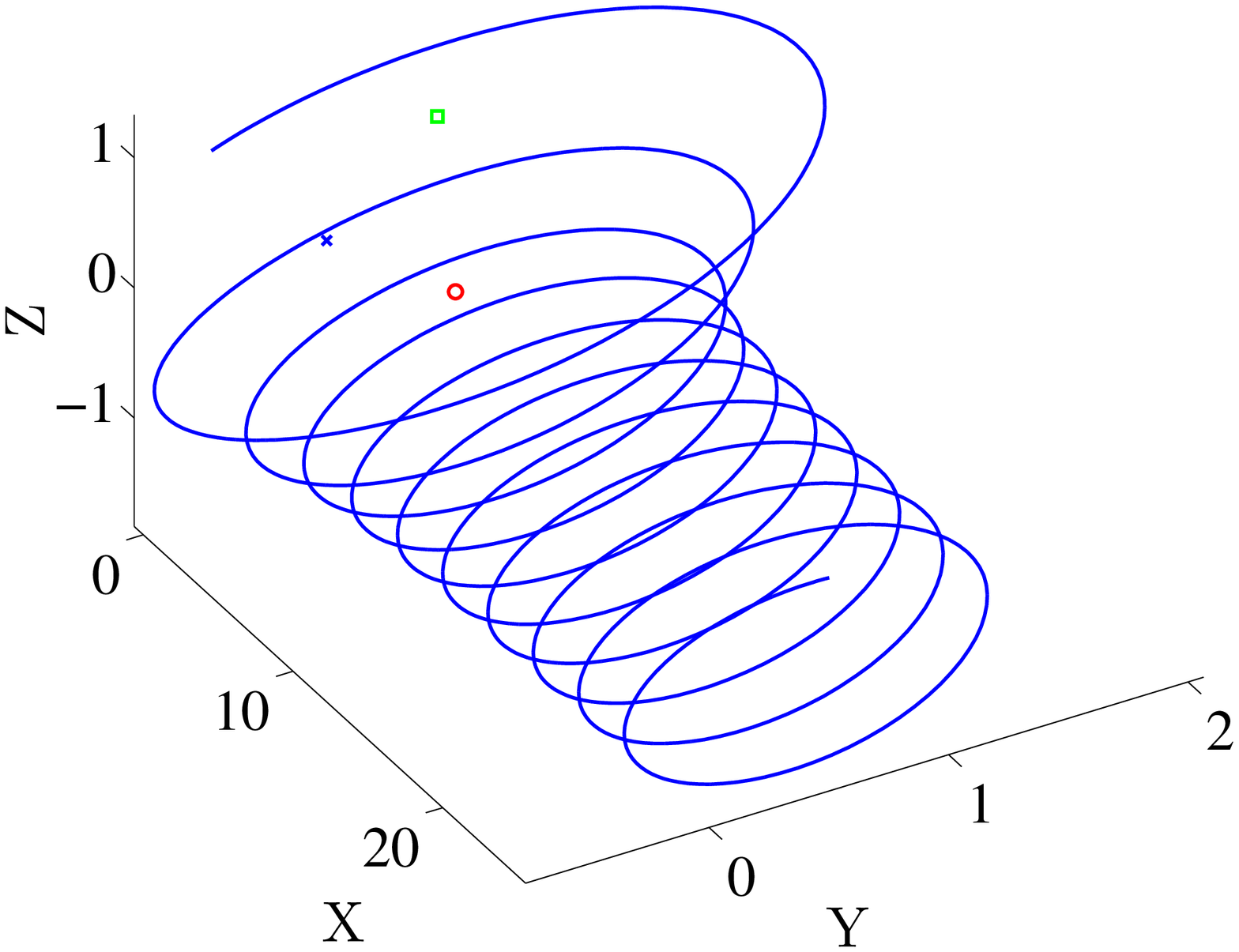}
\includegraphics[width=8cm]{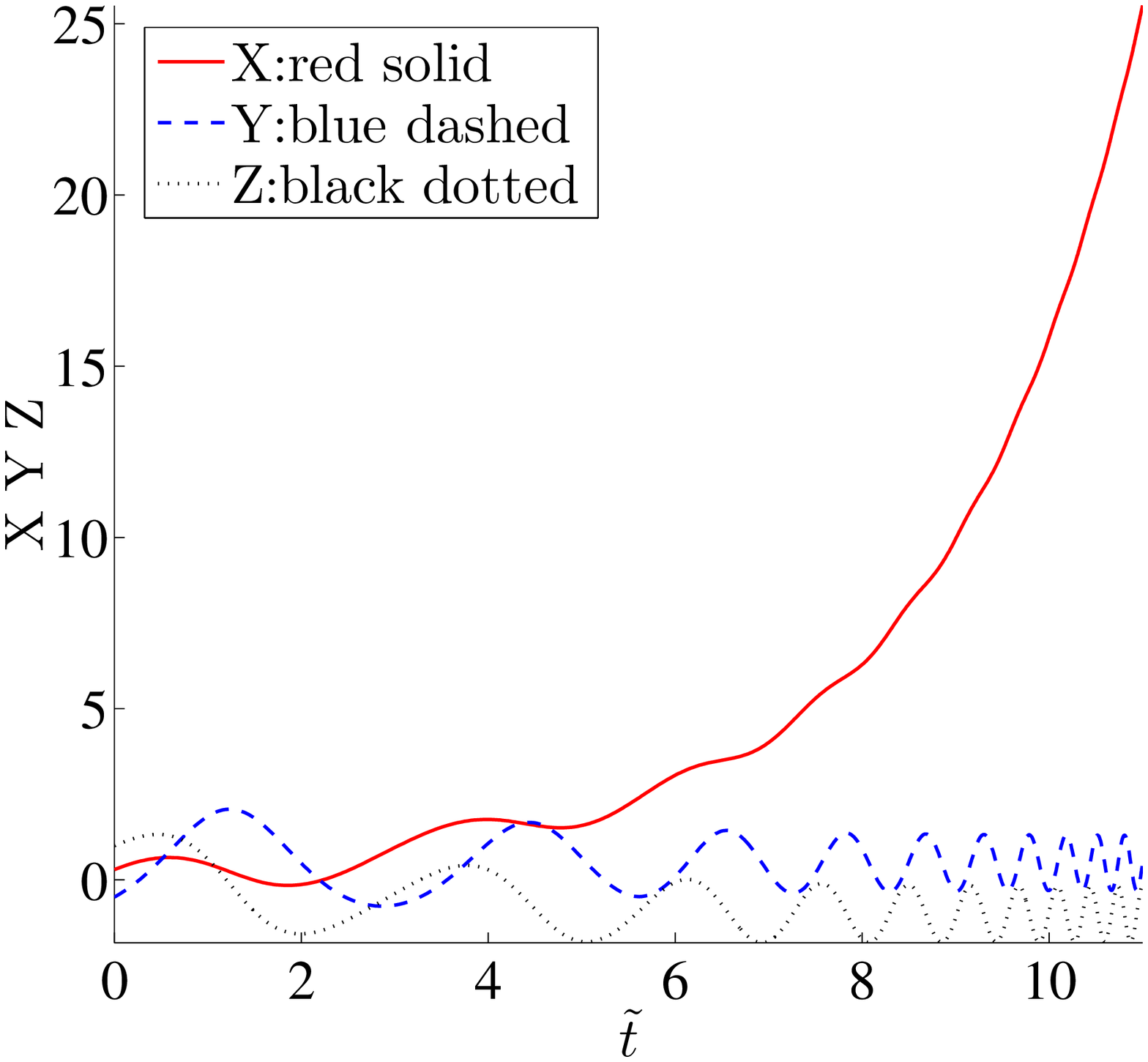}\\
\end{tabular}
\caption{Here, we only show (two examples of) the case with $\sigma^2 > b^2$.
In the top panels, $a=0.5$, $b=0.5$, $\sigma=1$, $P=1$, periodic-orbit-looking dynamics can
eventually escape to infinity becoming unbounded. For the bottom, if we move the initial condition even further, divergence shows up in the dynamics in a short time (red circle: $F^+$; blue cross: the origin; green square: $F^-$).}
\label{fig7_1}
\end{figure}

\begin{figure}[!htbp]
\begin{tabular}{cc}
\includegraphics[width=8cm]{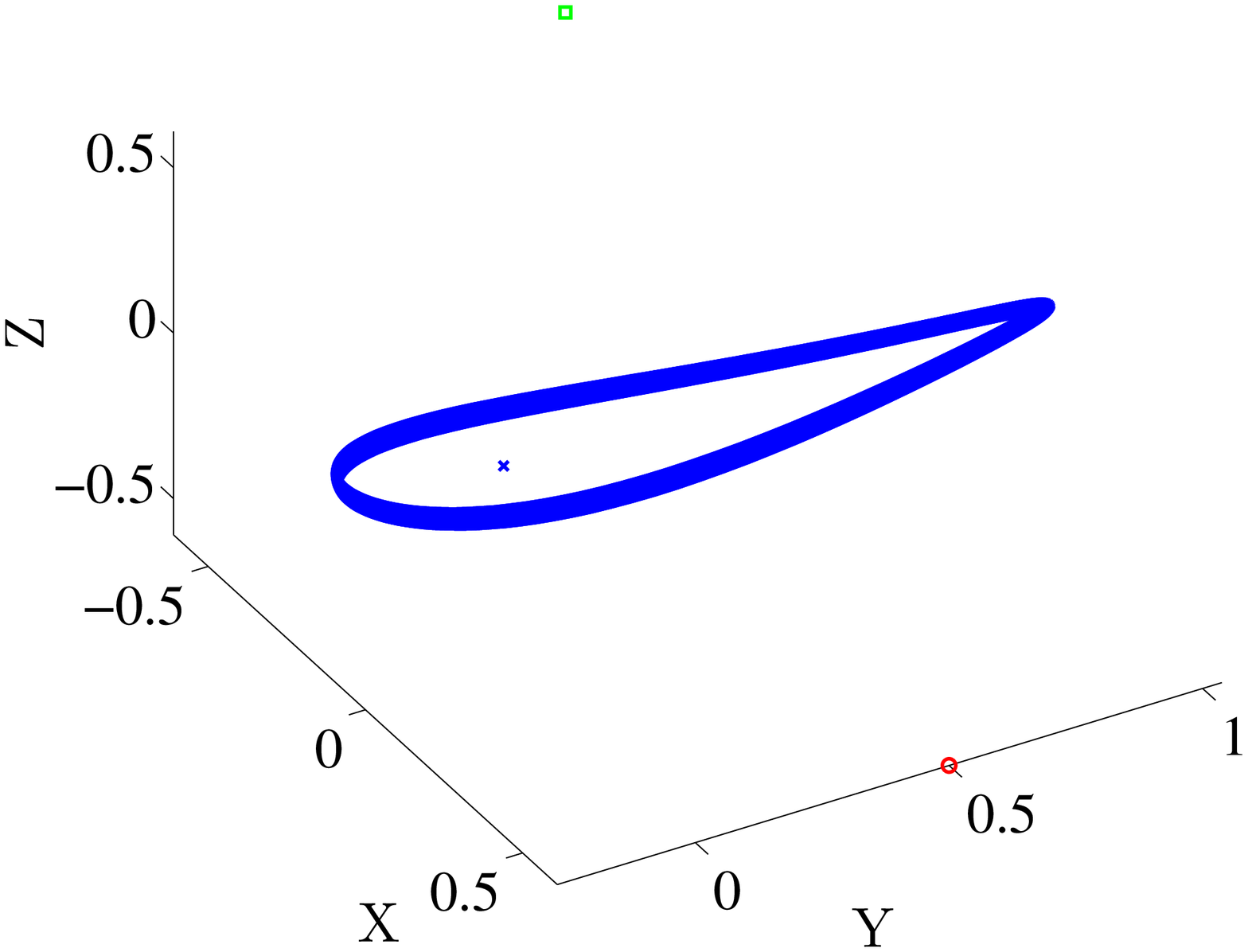}
\includegraphics[width=8cm]{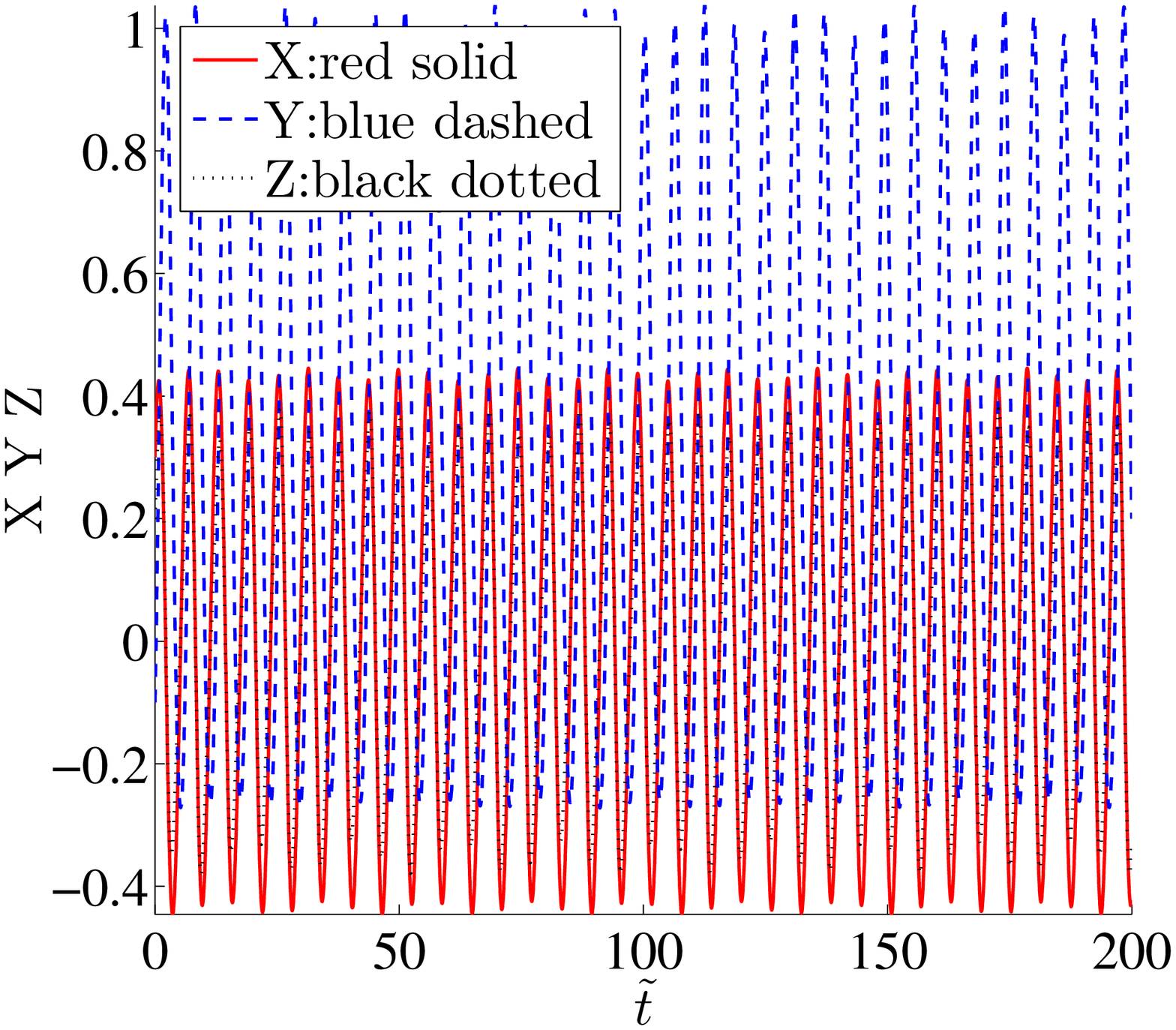} \\
\includegraphics[width=8cm]{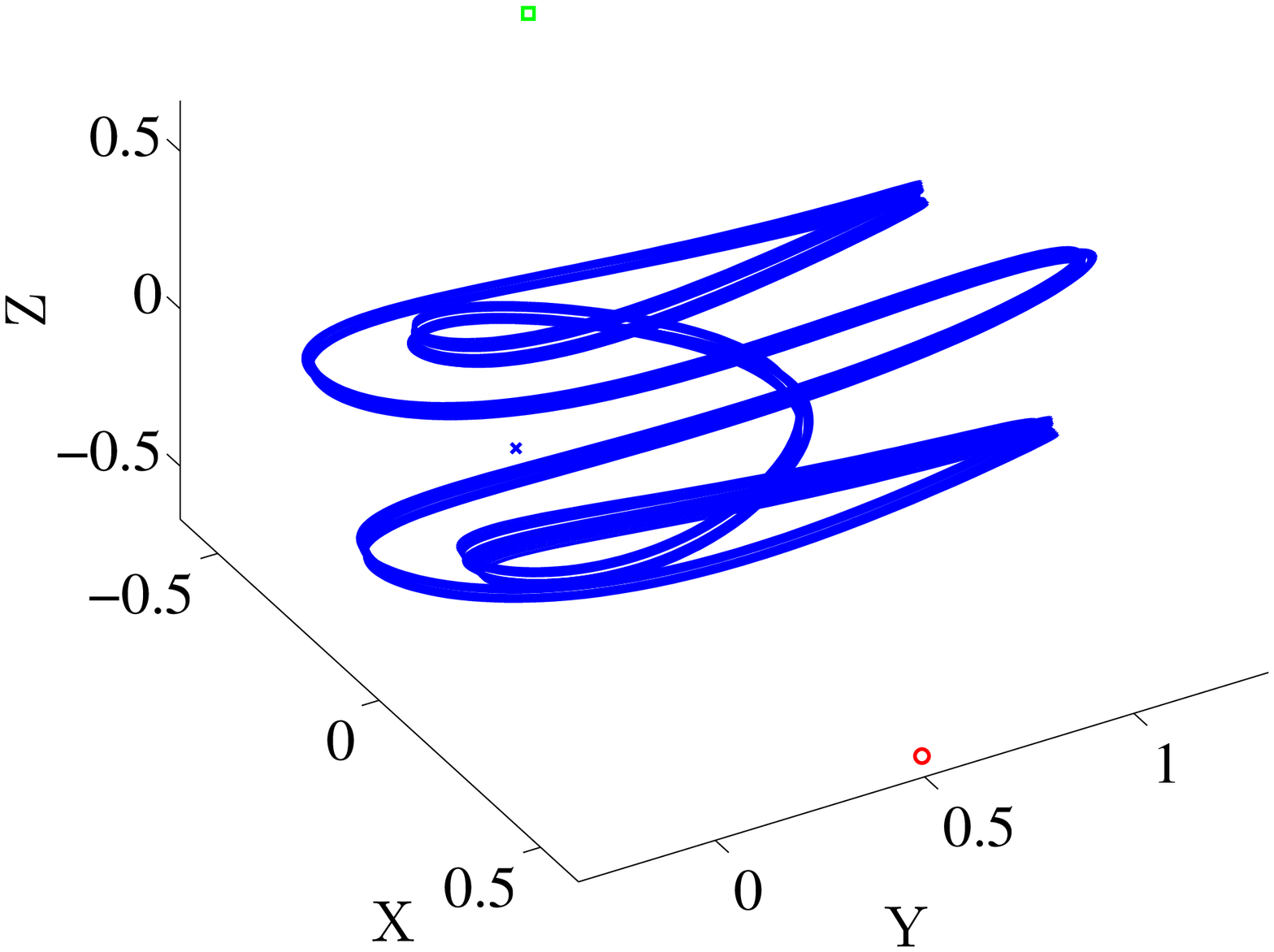}
\includegraphics[width=8cm]{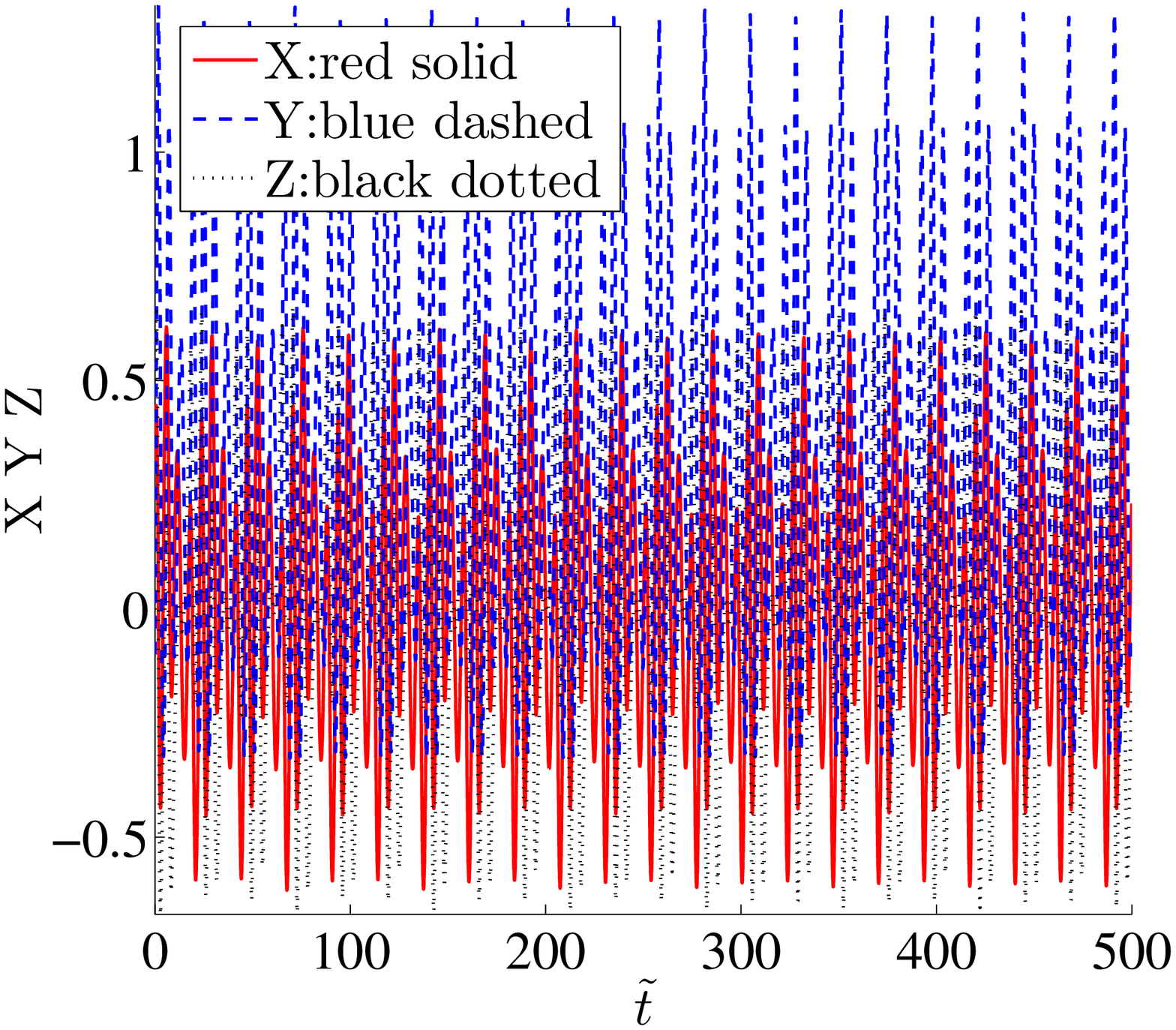} \\
\end{tabular}
\caption{Here we use same parameters as in Fig.~\ref{fig7_1} but different initial conditions.
In the top panels, we can observe here the existence of closed cycles in the $(X,Y,Z)$ plane that appear to correspond to
limit cycles in the dynamics. In the bottom panels, if we change the initial position a bit, closed orbits can still be obtained but they may look different
(red circle: $F^+$; blue cross: the origin; green square: $F^-$).}
\label{fig7_2}
\end{figure}

What is found in the dynamics, as well as from the detailed consideration
of different parameter values is that when $\sigma^2 < b^2$, the instability
of the origin appears to generically lead to divergence (whereby $X$ tends
to $\infty$). On the other hand, while the same behavior is possible
also for $\sigma^2 > b^2$ (see the bottom panels of
Fig.~\ref{fig7_1} for a relevant case
example), in the latter case it appears to also be possible
to observe periodic (closed cycle in the phase plane) behavior
around the fixed points. To be more specific, if we initialize somewhere between the fixed points $F^+$ and $F^-$, the system will just keep circling between $F^+$ and $F^-$, as illustrated in the two examples of Fig.~\ref{fig7_2}.
However, a slightly different initialization may lead to
wandering around these fixed points for a while
but eventually escaping to infinity~(see the top panels in Fig.~\ref{fig7_1}).
Notably, the result of the evolution in these cases appears to be
sensitively dependent
on the precise form of the initial condition.
Finally, choosing the initial position far from these fixed points leads
to a rapid growth of $|X|$ towards infinity.



\subsection{$a_1=-a_2=a$}

We now turn to a less physically relevant but still
mathematically interesting case, where the underlying
active medium leads to opposite effects between the
two waveguides. That is, it provides cross-coupling gain for one, while
it leads to corresponding loss for the other. In the following discussion, we show that this setup makes the system closely resemble the case $a=0$, which is a special case example of the setup of this section.

\subsubsection{$b_1=b_2=b$}

The system in this case features equal intrinsic gain (or loss)
between the waveguides. As Eq.~(\ref{eqn2_00}) suggests, we have $\frac{dN}{d\tilde{t}}=b N$ in this sub-case. Similar to case A.1, here the origin is the only possible fixed point of the system and it is either a sink~($b<0$) or a
source~($b>0$).

\subsubsection{$b_1=-b_2=b$, $P_1=P_2=P$}

We now turn to additional cases, such as in this section the
one where the intrinsic gain/loss pattern is anti-symmetric i.e.,
$b_1=-b_2$. Remarkably this case is $\mathcal{P T}$ symmetric
in its own right.

The system in the Stokes variables within this case reads:
\begin{eqnarray}
\frac{dX}{d\tilde{t}}&=&a Z-\sigma Y+ b N,\\
\frac{dY}{d\tilde{t}}&=&\sigma X+ P X Z,\\
\frac{dZ}{d\tilde{t}}&=&-a X-P X Y.
\label{eqn2i}
\end{eqnarray}
As a direct observation, for $b^2>a^2+\sigma^2$, in
the $\mathcal{P T}$-symmetry broken regime (see also the
discussion below), $|X|$ will increase or decrease monotonically.

Since subsection A.2 describes a special case of the current case, it should not
surprise us that techniques used in subsection A.2 can be applied here.
In fact, we can again write $Y$ and $Z$ into harmonic oscillator equations:
\begin{eqnarray}
Y(\tilde{t})&=&-\frac{a}{P}+C_1 \cos(P s(\tilde{t}))+C_2 \sin(P s(\tilde{t})),\\
Z(\tilde{t})&=&-\frac{\sigma}{P}+C_2 \cos(P s(\tilde{t}))-C_1 \sin(P s(\tilde{t})).
\label{eqn2i2}
\end{eqnarray}
where $C_1$ and $C_2$ are two constants related to the initial conditions. So $Y$ and $Z$ are always on a circle centered at $(-\frac{a}{P},-\frac{\sigma}{P})$ and $|u_1u_2|$ is bounded for all time $\tilde{t}$. Again, if we choose initial values of $u_1$ and $u_2$ carefully, either $|u_1|^2$ or $|u_2|^2$ will blow up to infinity, while the other will decay towards zero.

We find here that the origin is always a fixed point of the system, with the eigenvalues being $\pm \sqrt{ b^2- a^2 - \sigma^2}$, hence
the presence of the active medium {\it extends} the region of
$\mathcal{P T}$ symmetry in comparison to the well-known case of $a=0$. When $b^2>a^2+\sigma^2$ the origin is a saddle point. When $b^2<a^2+\sigma^2$, the origin becomes a center and fixed points $(0, \frac{b\sigma+a\sqrt{a^2+\sigma^2-b^2}}{\sigma^2+a^2}N, \frac{-ba+\sigma\sqrt{a^2+\sigma^2-b^2}}{\sigma^2+a^2}N)$, $(0, \frac{b\sigma-a\sqrt{a^2+\sigma^2-b^2}}{\sigma^2+a^2}N, \frac{-ba-\sigma\sqrt{a^2+\sigma^2-b^2}}{\sigma^2+a^2}N)$ (with $N$ nonnegative) arise. Their Jacobian matrices can be directly calculated as having
eigenvalues
$\lambda(\lambda^2-(b^2-a^2-\sigma^2\mp \sqrt{a^2+\sigma^2-b^2}PN))=0$.

As an alternative approach, we introduce cylindrical coordinates in order to
obtain a clearer picture of the dynamical system. Let $Y=-\frac{a}{P}+\rho\cos{\theta}$ and $Z=-\frac{\sigma}{P}+\rho\sin{\theta}$, then we obtain:
\begin{eqnarray}
\frac{d\rho}{d\tilde{t}}&=&0,\\
\frac{d\theta}{d\tilde{t}}&=&-P X,\\
\frac{d X}{d\tilde{t}}&=&a\rho\sin{(\theta)}-\sigma\rho\cos{(\theta)}+b\sqrt{(-\frac{a}{P}+\rho\cos{\theta})^2+(-\frac{\sigma}{P}+\rho\sin{\theta})^2}.
\label{eqn2i3}
\end{eqnarray}
Since $\frac{d N}{d\tilde{t}}=b X=-\frac{b}{P}\frac{d\theta}{d\tilde{t}}$, we can express $N$ using $\theta$ as $N=-\frac{b}{P}\theta+C$ where $C=N(0)+\frac{b}{P}\theta(0)$. Hence Eqn.~(\ref{eqn2i3}) can be rewritten as:
\begin{eqnarray}
\frac{d X}{d\tilde{t}}=a\rho\sin{(\theta)}-\sigma\rho\cos{(\theta)}-\frac{b^2}{P}\theta+b C.
\label{eqn2i4}
\end{eqnarray}
Note that the above equations are valid except at the origin or
when $\rho=0$. When $\rho=0$, the system is as simple as a straight line $Y=-\frac{a}{P}, Z=-\frac{\sigma}{P}$ and $X$ moves all the way to $\infty$~($b>0$)
or $-\infty$~($b<0$).

For other situations, once $\rho$ and $C$ are determined  by
the initial conditions, fixed points of the two-dimensional system satisfy:
\begin{eqnarray}
X=0, ~~~ a\rho\sin{(\theta)}-\sigma\rho\cos{(\theta)}-\frac{b^2}{P}\theta+b C=0.
\label{eqn2i5}
\end{eqnarray}
and corresponding eigenvalues are given by:
\begin{eqnarray}
\lambda^2=b^2-P\rho(a\cos{(\theta)}+\sigma\sin{(\theta)}).
\label{eqn2i6}
\end{eqnarray}


Inspired by forms of the dynamical equations of $X$ and $\theta$ and work related to the case $a=0$~\cite{barasflach2}, we state that there is an
effective energy conservation law for a classical particle
fully describing the system:
\begin{eqnarray}
\frac{1}{2}\left(\frac{dq}{d\tilde{t}}\right)^2+V(q)=E ,
\label{eqn2i7}
\end{eqnarray}
with displacement $q=\theta-\frac{P}{b}C$, potential $V(q)=-\frac{b}{2}q^2-aP\rho\cos{(q+\frac{P}{b}C)}-\sigma P\rho\sin{(q+\frac{P}{b}C)}$ and energy $E=\frac{1}{2}P^2\rho^2-\frac{1}{2}a^2-\frac{1}{2}\sigma^2$. Exploring the dynamics and phase portrait of this classical system can help us explain our original system, and yields results quite similar to the case A.2. Again, the origin and case $\rho=0$ are excluded from Eqn.~(\ref{eqn2i7}).

The number of fixed points on the cylinder varies as the radius $\rho$ changes, as it is clearly seen from the expressions of fixed points including $N$. When the radius $\rho$ is small such that $\rho<|\frac{b}{P}|$, there are
no fixed points on the cylinder (except the origin as a saddle with $\rho=\sqrt{\frac{a^2+\sigma^2}{P^2}}<|\frac{b}{P}|$) and all trajectories will
become unbounded. As $\rho$ grows beyond $|\frac{b}{P}|$, two
additional fixed points arise on the cylindrical surface. It can be checked via their eigenvalues that one of them is center and the other is a saddle.
Thus some trajectories can form periodic orbits around the center and others simply run away to infinity along the cylinder.

\subsubsection{$b_1=-b_2=b$, $P_1=-P_2=P$}

Finally, we examine this case, which is interesting in its own right,
as it is $\mathcal{P T}$-symmetric at the linear level, while the
nonlinearity breaks the symmetry.
The system in this case is:
\begin{eqnarray}
\frac{dX}{d\tilde{t}}&=&a Z-\sigma Y+ b N,\\
\frac{dY}{d\tilde{t}}&=&\sigma X+ P N Z,\\
\frac{dZ}{d\tilde{t}}&=&-a X-P N Y.
\label{eqn2k}
\end{eqnarray}
As usual, in this case too, the origin is always a fixed point and actually it is the only one.
Its stability is determined by the same conditions
as in case C.2. Furthermore, when $b^2>a^2+\sigma^2$, $\frac{dX}{d\tilde{t}}$ will never change its sign over time, as is shown in the previous case.

As a general version of case A.3, we can similarly construct a special function $L_1=N X-\frac{a}{P}Y-\frac{\sigma}{P}Z$. Then its dynamical equation reads $\frac{d L_1}{d\tilde{t}}=b(N^2+X^2)$ so that $L_1$ always increases or decreases. This property of $L_1$ disproves (similarly
to the argument discussed in the case A.3)
the potential existence of any periodic orbits in the system.

\begin{figure}[!htbp]
\begin{tabular}{cc}
\includegraphics[width=8cm]{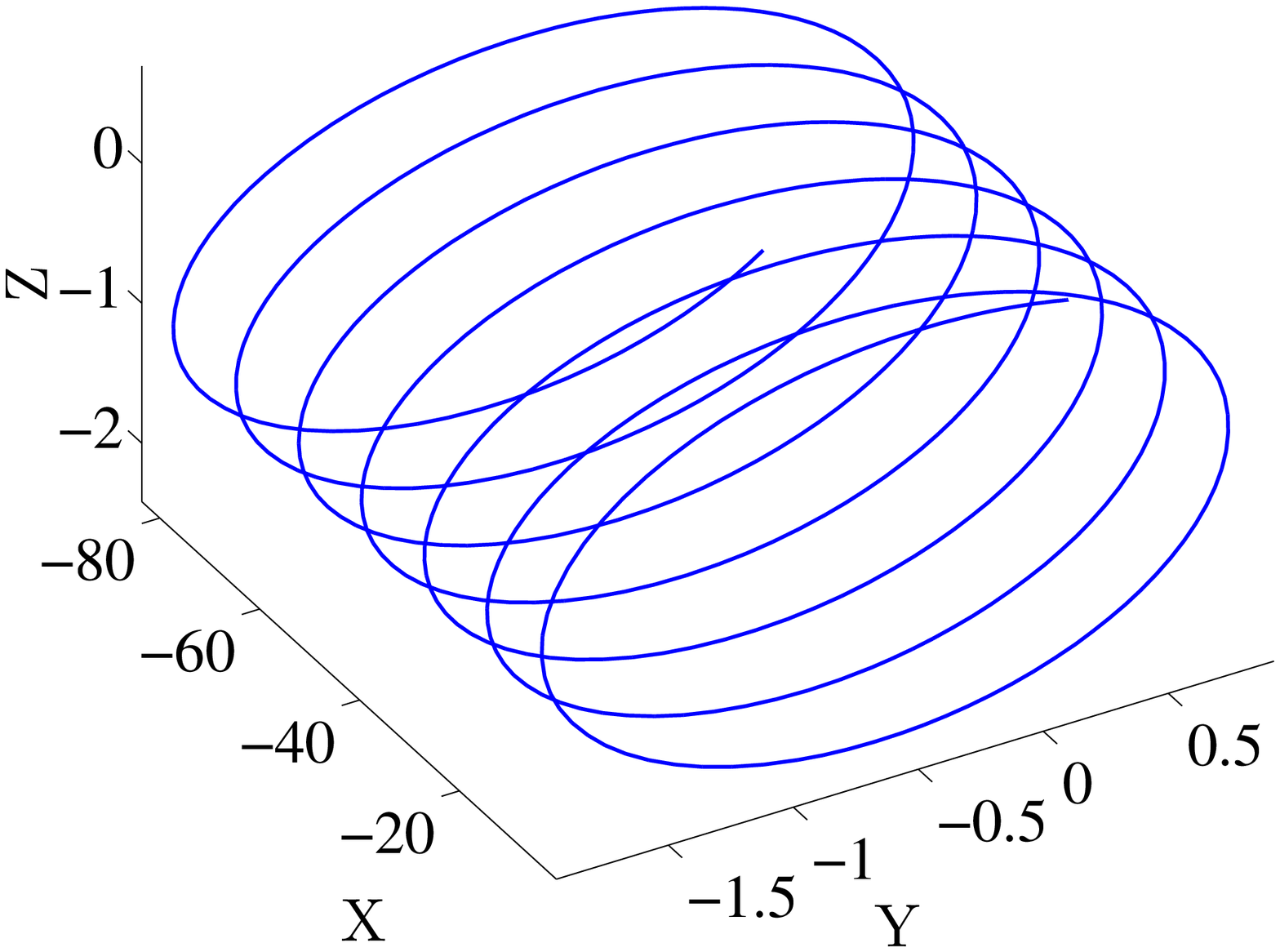}
\includegraphics[width=8cm]{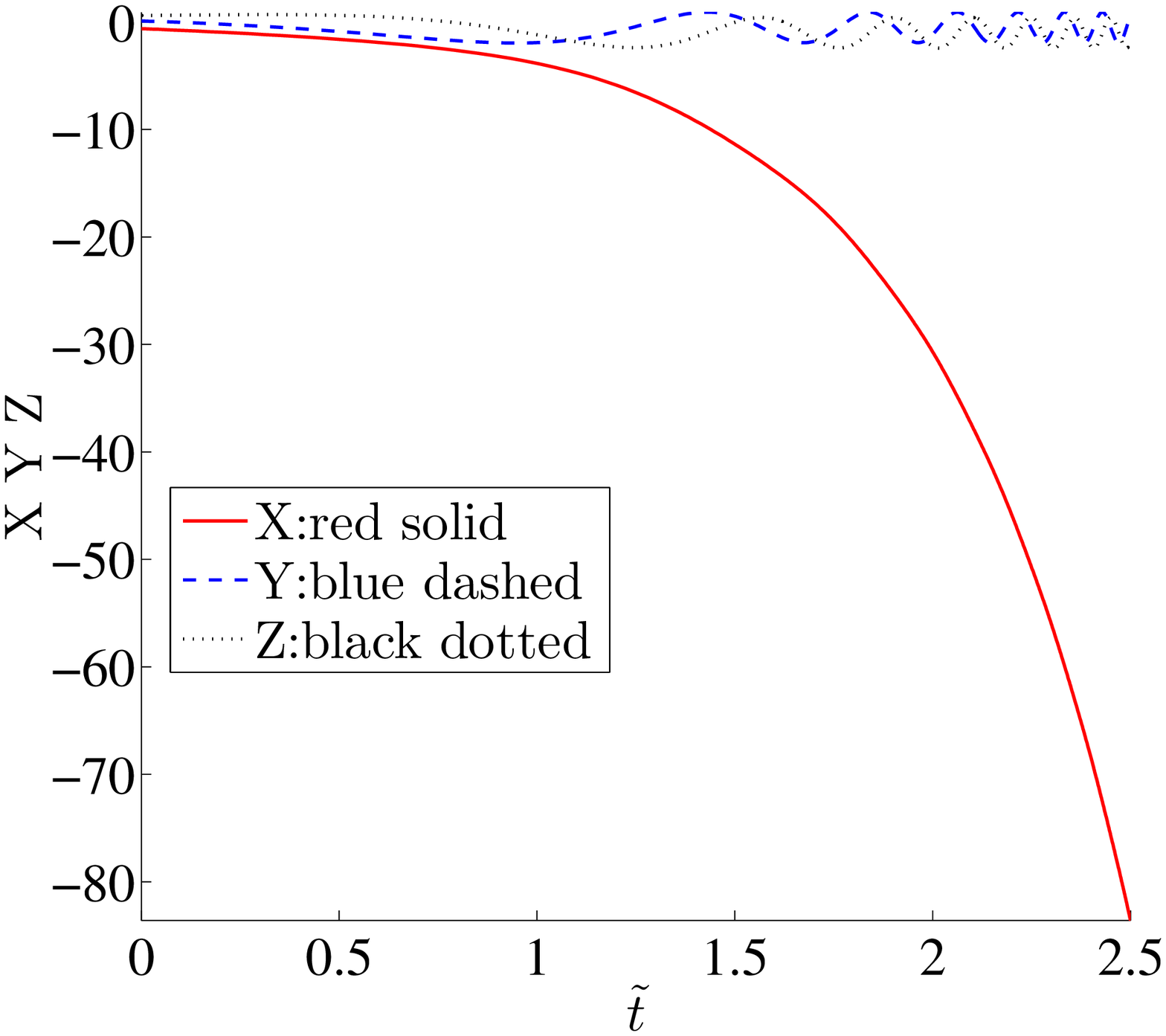} \\
\includegraphics[width=8cm]{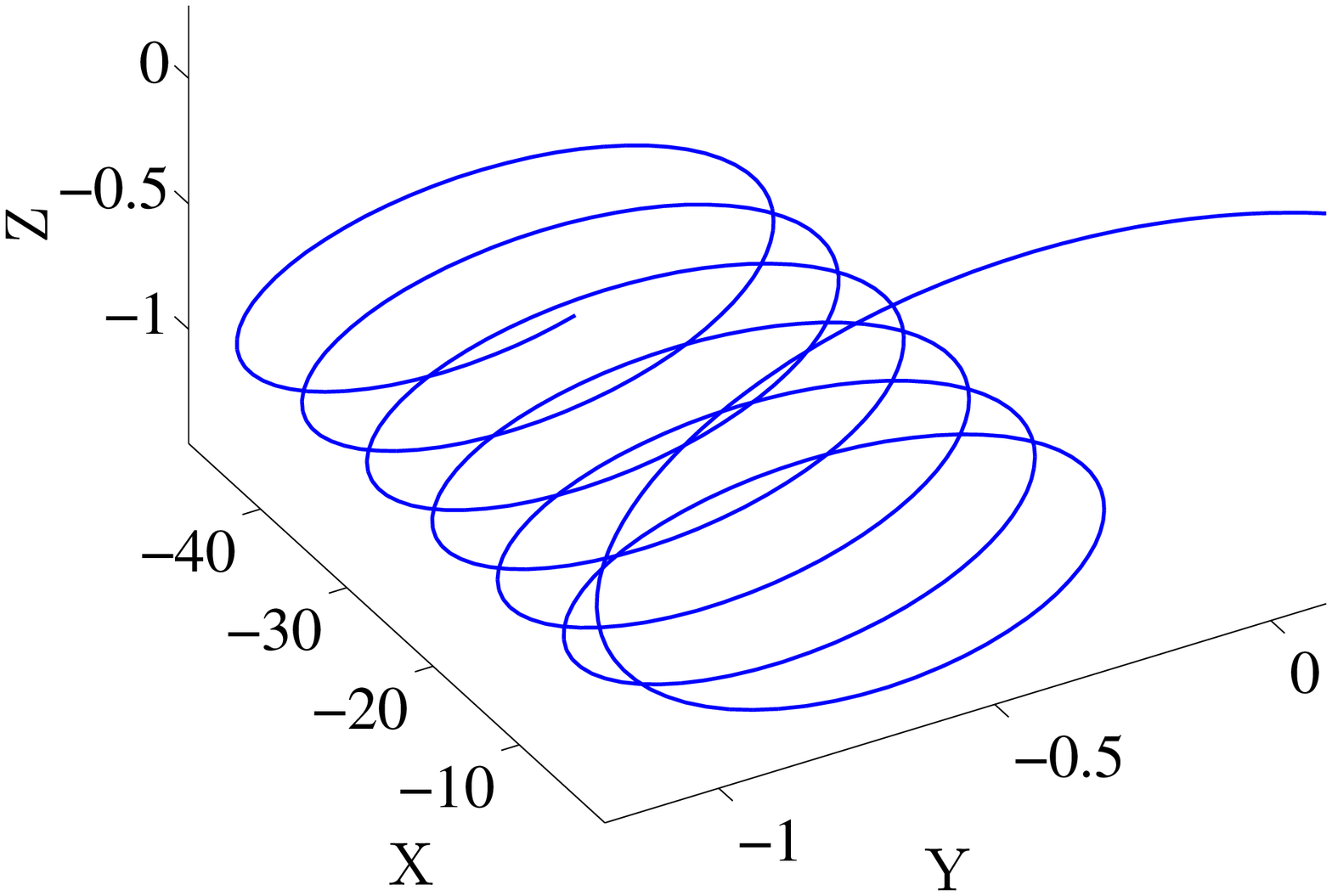}
\includegraphics[width=8cm]{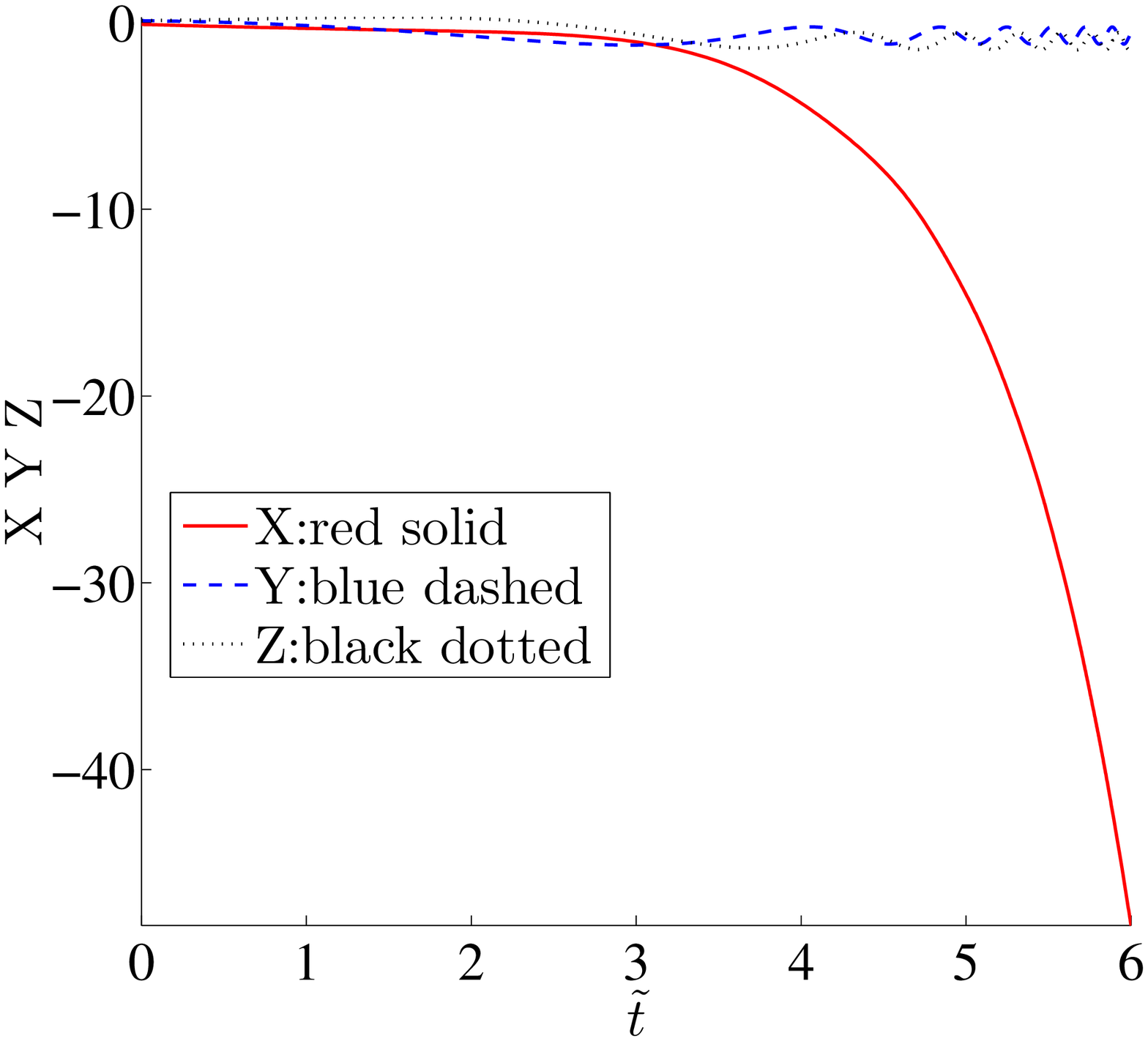}
\end{tabular}
\caption{The top panels are under the condition $b^2>a^2+\sigma^2$ with $a=0.5$, $b=-2$, $\sigma=1$, $P=-1$, showing the origin $(0,0,0)$ is not stable here since $X\to-\infty$. The bottom panels
are for the case of
$b^2<a^2+\sigma^2$ and $a=0.7$, $b=-1.2$, $\sigma=1$, $P=-1$. Again, the evolution will lead to the blowing up of $|X|$ instead of circling around the origin as in case C.2.}
\label{fig8}
\end{figure}

For the dynamics in Stokes variables, we find that
it repeats some features reported
in case A.3. When $b^2>a^2+\sigma^2$, as shown from the dynamical equation of $X$, $X$ clearly goes to $\infty$~($b>0$) or $-\infty$~($b<0$);
see top panels in Fig.~\ref{fig8}. Even when
$b^2 < a^2+\sigma^2$, the evolution still leads to unboundedness,
especially for $X$, as captured in the bottom panels of Fig.~\ref{fig8}.

\section{Conclusions and Future Challenges}

In the present work, we have explored a series of generalized
examples of the dimer system. Our considerations involved
a number of special cases such as the ``standard'' $\mathcal{P T}$
symmetric dimer~\cite{kot1,sukh1,pgk,R44,R46,tyugin,pickton},
the actively coupled optical waveguides of~\cite{barasflach} for
particular values of the parameters, as well as numerous cases that
have not been previously considered, such as a $\mathcal{P T}$ symmetric
variation of the actively coupled system or another variant which is
$\mathcal{P T}$ symmetric at the linear level but nonlinearity
destroys the symmetry. Generally, we introduced and examined
the possibility of not only having nonlinearly uniform waveguides
but also waveguides constructed of different materials and hence
with different (here considered as opposite) Kerr coefficients.

In general, this broad class of systems led to a wide range of
interesting behaviors and bifurcation phenomena. In addition to
analyzing the stability of the origin, we identified pitchfork
bifurcations that gave rise to new fixed points, and Hopf bifurcations
that led to the emergence (e.g. in the active medium case) of limit
cycles.
Moreover, chaotic dynamics
was revealed in some of the cases. Furthermore, various diagnostics
including the use of Stokes variables and of their dynamical
systems' analysis were brought
to bear in order to characterize the dynamics of the different
subcases.

Naturally, there are numerous veins for the extension of the
present study. On the one hand, for one dimensional systems,
it is quite relevant to extend these dimer settings to
other oligomer cases, including examples of trimers~\cite{pgk,kaiser},
quadrimers~\cite{pgk,konorecent3} and appreciate how the phenomenology
observed herein is extended in this larger number of degrees of freedom
cases. On the other hand, extending the relevant phenomenology to higher
dimensions and plaquette type configurations as in the case of~\cite{uwe}
would be another natural possibility in its own right. Such possibilities
are currently under active consideration and will be reported in
future publications.

\end{document}